\documentclass[final,onefignum,onetabnum]{siamart171218}

\usepackage{amssymb, amsfonts, amsmath}

\title{A local continuum model of cell-cell adhesion\thanks{Submitted to the editors 28/06/22
\funding{JAC was supported by the Advanced Grant Nonlocal-CPD (Nonlocal PDEs for Complex Particle Dynamics: Phase Transitions, Patterns and Synchronization) of the European Research Council Executive Agency (ERC) under the European Union’s Horizon 2020 research and innovation programme (grant agreement No. 883363).
JAC was also partially supported by EPSRC grants EP/T022132/1 and EP/V051121/1. CF acknowledges support of a fellowship from "la Caixa" Foundation (ID 100010434) with code LCF/BQ/EU21/11890128.}}}

\author{C. Falc\'o, R. E. Baker, J. A. Carrillo\thanks{Mathematical Institute, University of Oxford, OX2 6GG Oxford, United Kingdom} }

\begin{document}

\maketitle

\begin{abstract}
Cell-cell adhesion is one the most fundamental mechanisms regulating collective cell migration during tissue development, homeostasis and repair, allowing cell populations to self-organize and eventually form and maintain complex tissue shapes. Cells interact with each other via the formation of protrusions or filopodia and they adhere to other cells through binding of cell surface proteins. The resulting adhesive forces are then related to cell size and shape and, often, continuum models represent them by nonlocal attractive interactions. In this paper, we present a new continuum model of cell-cell adhesion which can be derived from a general nonlocal model in the limit of short-range interactions. This new model is local, resembling a system of thin-film type equations, with the various model parameters playing the role of surface tensions between different cell populations. Numerical simulations in one and two dimensions reveal that the local model maintains the diversity of cell sorting patterns observed both in experiments and in previously used nonlocal models. In addition, it also has the advantage of having explicit stationary solutions, which provides a direct link between the model parameters and the differential adhesion hypothesis.
\end{abstract}



\section{Introduction}

From the formulation of the differential adhesion hypothesis (DAH) by Malcolm Steinberg more than 50 years ago, there have been many experimental and modeling efforts  to understand adhesion-based cellular self-organization. Differential adhesion between cell populations is now understood as a fundamental mechanism for the formation of tissue shapes during morphogenesis, maintenance and repair, as it allows cells to sort and arrange themselves into complex patterns. On the other hand, mathematical models of differential adhesion not only provide helpful insights into experimental work but have also proven to be interesting from an analytical point of view, motivating a number of further theoretical studies.

The experimental evidence of adhesion-based sorting provided by Townes and Holtfreter \cite{holtfreter1943properties,townes1955directed} inspired Steinberg to formulate the DAH \cite{steinberg1962II,steinberg1962I,steinberg1962III,steinberg1963reconstruction}.  These experiments showed that dissociated cell populations from amphibian embryos could self-organize and arrange themselves into a pattern with two distinguished cell types. Based on the analogy that cells behave as immiscible fluids with different surface tensions, just like oil and water, Steinberg developed a thermodynamic model that could explain this behaviour in terms of the relative strengths of cell-cell adhesion bonds.   Under this setting then, when different cell populations are mixed, they self-organize in order to minimize the total adhesion energy of the tissue. This framework has been able to explain for instance the bullseye pattern seen in multiple experiments \cite{duguay2003cadherin,foty2005differential,krieg2008tensile}, in which two cell populations sort into two concentric spheroids, with the most adherent cells comprising the inner one. The DAH and further experimental evidence supporting it are reviewed in \cite{foty2004cadherin} -- see also  \cite{reviewCellCellAdhesion} for a more modern perspective.

In this paper, we present a local continuum model of cell-cell adhesion, in which differential surface tension between cell populations is able to reproduce the diversity of patterns described by the DAH. This new model
is based on a system of thin-film type equations, and while it could be considered as a phenomenological model, it can be  derived from a general nonlocal model in the limit of short-range interactions.
In its reduced version, the model only has four parameters which admit a physical interpretation both in the context of nonlocal models, and in terms of surface tensions for thin-film equations \cite{myers1998thin}. Further, the local model has the advantage of being more analytically tractable than nonlocal models, as it presents explicit stationary solutions even in the case of two interacting species.

\raggedbottom

\subsection{Nonlocal models of cell-cell adhesion}

Mathematical models describing  cell-cell adhesion have taken different approaches by considering either interfacial energy contributions, a tissue bulk modulus, or short-range attraction in the form of nonlocal interactions \cite{AlertTrepat}. Many individual-based models have been used for adhesion-based patterning, e.g. cellular Potts type models \cite{hirashima2017cellular,krieg2008tensile}, vertex models \cite{VertexModels,hashimoto2017}, and particle-based models \cite{CarrilloColombiScianna,VolkeningZebrafish}, to name but a few. While these have been successful in reproducing the observed experimental patterns, discrete models present difficulties, namely the computational cost involved in solving them and the lack of analytic insights for large numbers of cells. Continuum models, in principle, can offer a solution to these issues, but it was not until a decade ago that this was achieved in the context of cell-cell adhesion.

The first continuum model of cell-cell adhesion able to reproduce cell sorting phenomena was initially proposed by \cite{ArmstrongPainterSherratt}, and is based on the idea that cells move according to random motion, which results in linear diffusion, and cell-cell adhesion. The latter is represented by a nonlocal attractive term which emerges from assuming adhesive forces between cells that are within a given distance or \emph{sensing radius}. One caveat resulting from the random motion assumption however, is that in some situations, the model shows unrealistic biologic behaviour. For instance, it does not predict full seggregation nor sharp boundaries. In order to mitigate this issue, the linear diffusion term may be substituted by a density-dependent diffusion term accounting for population pressure \cite{MurakawaTogashi}. Such nonlinear diffusion equations are often used to describe crowding effects in mathematical biology \cite{calvezCarrillo,DysonVolumeExclusion,Gurtin1977OnDiffusion} and can be derived from individual-based models \cite{DysonMacroscopicCrowding}, as well as from on-lattice models \cite{bakerAspectRation,falco2022random}. The modified model and variations of it with density-limited mobilities  \cite{CarrilloMurakawaCellAdhesion} have proven to show a more accurate behaviour of adhesion-based pattern formation.

For the sake of conciseness here we do not explicitly derive the mentioned nonlocal models, but we refer to \cite{chenpainter2020nonlocal} for further and more detailed explanations. However, a fairly general nonlocal model related to the ones above can be derived as the thermodynamic limit of a system of interacting particles \cite{CarrilloMurakawaCellAdhesion}. In this model, cells interact with other cells via a strong repulsion at short distances due to the volume-filling effect of the cell nuclei, and by attraction at medium distances. The latter is linked to the size of the cell and its protrusions or filopodia, and represents adhesive forces.

Consider then a system of $N$ interacting cells whose positions are given by $\{\mathbf{y}_i\}$ for $i = 1,\ldots,N$. For simplicity, we assume now there is only one cell population and hence the forces exerted between cells can be modeled as the gradient of a given potential $W^N$, which in the case of isotropic interactions is radial. The basic individual-based model for this system reads
\begin{equation*}
    \frac{\mathrm{d}\mathbf{y}_i}{\mathrm{d}t} = -\frac{1}{N}\sum_{j\neq i}\nabla W^N\left(\mathbf{y}_i-\mathbf{y}_j\right),\quad \mbox{for }i = 1,\ldots,N.
\end{equation*}
In the limit of large $N$ one is interested in describing cell density $\rho(\mathbf{x},t)$, $\mathbf{x}\in\mathbb{R}^d$, instead of individual cell trajectories.
For that purpose, we define the so-called empirical measure
\begin{equation*}
    \rho^N(\mathbf{x},t) = \frac{1}{N}\sum_{i = 1}^N\delta_{\mathbf{y}_i(t)},
\end{equation*}
where $\delta_{\mathbf{y}_i(t)}$ is a Dirac delta measure centered at $\mathbf{y}_i(t)$.

We now take into account the specific shape of the potential $W^N$ and how it scales with the number of cells. One way to represent volume exclusion is to assume that for small distances,  $W^N$ approaches a Dirac delta, $\delta_0$, at the origin as $N\rightarrow\infty$.
This scaling has been studied rigorously in \cite{Oelschlger1990LargeSO}, where the following form of the potential is considered
\begin{equation*}
    W^N(\textbf{x}) = \epsilon N^{\beta}\psi\left(N^{\beta/d}\mathbf{x}\right)
 + W(\textbf{x}),
\end{equation*}
with $\psi$ being a typical repulsive potential with unit volume, and $W$ a purely attractive potential. The parameter $\epsilon>0$ measures the relative strength of repulsion to attraction. Under this scaling, and for any $\beta\in(0,1)$, the empirical measure in the limit $N\rightarrow\infty$ can be identified with the solution of the aggregation-diffusion equation
\begin{equation}
\label{eq:nonlocal_1species}
    \frac{\partial\rho}{\partial t} = \nabla\cdot\left(\rho\nabla\left(\epsilon\rho + W*\rho\right)\right),
\end{equation}
with $(W*\rho)(\mathbf{x},t) = \int_\Omega W\left(\mathbf{x}-\mathbf{y}\right)\rho(\mathbf{y},t)\mathrm{d}\mathbf{y}$.
The above model is closely related to the ones in \cite{ArmstrongPainterSherratt,MurakawaTogashi}.

In the case of two interacting species \cite{CarrilloMurakawaCellAdhesion}, given by $\rho$ and $\eta$, one can follow the same ideas to obtain
\begin{subequations}
\begin{align}
    \frac{\partial \rho}{\partial t} = \nabla\cdot\left(\rho\nabla \left(W_{11}*\rho + W_{12}*\eta +\epsilon(\rho+\eta) \right)\right), \label{eq:nonlocal_2species_a} \\
    \frac{\partial \eta}{\partial t} = \nabla\cdot\left(\eta\nabla \left(W_{21}*\rho + W_{22}*\eta +\epsilon(\rho+\eta) \right)\right),\label{eq:nonlocal_2species_b}
\end{align}
\label{eq:nonlocal_2species}
\end{subequations}
where $W_{11},\,W_{22}$ are the self-adhesion potentials, and $W_{12},\,W_{21}$ represent the cross-adhesion interactions. Again, the parameter $\epsilon>0$ measures the strength of the localized repulsion. Existence of solutions for this system is proven in \cite{AntonioTwoSpeciesNonlocal} -- see also \cite{difrancesco2013measure} for the case without cross-diffusion. A popular choice is to assume that the cross-interaction is symmetrical $W_{12} = W_{21}$, and that the potentials have the same shape $W_{ij} = K_{ij} W$, with $W$ a typical attractive potential, and the constants $K_{ij}\geq 0$ giving the cell-cell adhesion strengths. This assumption on the shape of the potentials is related to previous nonlocal models of cell-cell adhesion \cite{ArmstrongPainterSherratt,CarrilloMurakawaCellAdhesion,MurakawaTogashi}.

\subsection{Outline}

Here, we follow the approach in \cite{BernoffTopazCH}, and derive a local model of cell-cell adhesion from Eqs. \eqref{eq:nonlocal_2species}. This model can be formally identified as an approximation in the limit of short-range interactions -- or as a long-wave approximation. However, the goal of the paper is not to compare these local and nonlocal models, but to study the former, and explore if it is consistent with the DAH.

Upon formally taking the limit of short-range interactions in the general nonlocal model given by Eqs \eqref{eq:nonlocal_2species},  we obtain a system of thin-film like equations modelling the evolution of the two cell populations
\begin{subequations}
\begin{align}
     \frac{\partial \rho}{\partial t}& = -\nabla\cdot\left(\rho\nabla \left(\kappa\Delta \rho + \alpha\Delta \eta + \mu\rho + \omega\eta \right)\right);\label{eq:intro1}\\
     \frac{\partial \eta}{\partial t}& = -\nabla\cdot\left(\eta\nabla \left(\alpha\Delta \rho + \Delta \eta + \omega\rho + \eta \right)\right)\label{eq:intro2}.
\end{align}
\label{eq:local_2species_intro}
\end{subequations}
The parameters in the system, $\kappa,\alpha,\mu\geq 0,\,\omega\in\mathbb{R}$, can be related to the potentials of the nonlocal model, $W_{ij}$, and to the strength of the volume-filling mechanism, but can also be understood as relative surface tensions, as in the thin-film equation. In this setting, one could ask whether differential tension -- analogous to differential adhesion -- in the model is sufficient to give rise to the  patterns seen in Steinberg experiments. Interestingly, we show that it is possible to identify parameter regimes for each one of the different observed configurations with the cross-interaction parameters $\alpha$ and $\omega$ playing a major role in the behaviour of the local model (see Figure \ref{fig:4patterns}).

This paper is structured in two parts. First we derive and study the local model for one cell population, including linear stability, numerical simulations of the model in one and two dimensions, and the calculation of steady states and associated energy minimizers. Then, we extend these ideas and derive the model for two interacting cell populations, Eqs. \eqref{eq:local_2species_intro}.
We show via numerical simulations that we can recover the patterns predicted by the DAH. Again, in the local model for two species, explicit stationary solutions are available and offer a direct interpretation for cell sorting phenomena.
Finally, we summarize our findings and discuss some other advantages of the local model compared to previously used nonlocal models.

\section{One species model}\label{sec:one_species}
\subsection{Heuristic derivation of the model and basic properties}

We begin with the nonlocal model given in Eq.  \eqref{eq:nonlocal_1species}. Recall that the $\epsilon\rho$ term represents a localized repulsive force at the origin and the potential $W$ is assumed to be purely attractive and radially symmetric.

Current models of adhesion only take into account interactions between cells that are separated by less than a maximum \emph{sensing radius}. Here, we build on the idea that for large populations, such \emph{sensing radius} is much smaller than the typical size of the population and hence attractive forces between cells are given by a short-range interaction potential. Hence, we set $W(\mathbf{x}) =- a^{-d}\varphi(\mathbf{x}/a)$ with $a$ a scaling parameter which dictates the range of attraction, and $\varphi$ a sufficiently smooth function defined in $\mathbb{R}^d$. As  $a\rightarrow 0$, the potential $W$ tends to a Dirac delta function supported at the origin.
We further assume that the function $\varphi$ satisfies several conditions.
\begin{enumerate}
    \item $\varphi(\mathbf{x}) = \varphi(|\mathbf{x}|)$ and $\varphi(r)$ is a non-increasing function for $r>0$, meaning that $W$ is both symmetric and attractive.
    \item $\varphi$ approaches a constant as $r\rightarrow\infty$. Without loss of generality we assume that this constant is zero.
    \item The moments   $M_{n} = \int_{\mathbb{R}^d}|\mathbf{x}|^n\varphi(\mathbf{x})\,\mathrm{d}\mathbf{x}$ decay suitably fast.

\end{enumerate}
Omitting the time dependence and writing $(W*\rho)(\mathbf{x}) = -\int_{\mathbb{R}^d}\varphi(\mathbf{y})\rho(\mathbf{x}-a\mathbf{y})\mathrm{d}\mathbf{y}$, we can now consider the limit of short-range attraction and expand $\rho(\mathbf{x}-a\mathbf{y})$ as a Taylor series for small values of the scaling parameter $a$:
\begin{align*}
     (W*\rho)(\mathbf{x}) =&  -\rho(\textbf{x})\int_{\mathbb{R}^d}\varphi(\textbf{y})\,\mathrm{d}\textbf{y}
          + a\int_{\mathbb{R}^d}\left(\nabla\rho(\textbf{x})\cdot \textbf{y}\right)\varphi(\textbf{y})\,\mathrm{d}\textbf{y}
    \\& -\frac{a^2}{2} \int_{\mathbb{R}^d}\left(\textbf{y}^t\cdot H_\rho(\textbf{x}) \textbf{y}\right)\varphi(\textbf{y})\,\mathrm{d}\textbf{y}+o(a^2);
\end{align*}
where $H_\rho(\textbf{x})$ is the Hessian matrix of $\rho$.

We will only keep the first terms in the expansion. For the first term in the Taylor expansion we simply have
    $\rho\int_{\mathbb{R}^d}\varphi = M_0\rho$, and we also note that the terms with odd order derivatives of $\rho$ vanish due to the symmetry assumption on the potential. Then the error term in the expression above is  $O(a^4)$.
    The next non-vanishing term in the series contains the second-order derivatives of $\rho$ and reads
    \begin{align*}
 \int_{\mathbb{R}^d}\left(\textbf{y}^t\cdot H_\rho(\textbf{x}) \textbf{y}\right)\varphi(\textbf{y})\,\mathrm{d}\textbf{y}&=\sum_{i = 1}^d\sum_{j = 1}^d\frac{\partial^2\rho}{\partial x_i\partial x_j}\int_{\mathbb{R}^d}y_iy_j\varphi(\mathbf{y}) \,\mathrm{d}\mathbf{y}\\& =   \sum_{i=1}^d\frac{\partial^2\rho}{\partial x_i^2}\int_{\mathbb{R}^d}y_i^2\varphi(\mathbf{y}) \,\mathrm{d}\mathbf{y}\\& = \frac{1}{d} \left(\int_{\mathbb{R}^d}|\mathbf{y}|^2\varphi(\mathbf{y}) \,\mathrm{d}\mathbf{y}\right)\sum_{i=1}^d\frac{\partial^2\rho}{\partial x_i^2}=\frac{M_2}{d}\Delta \rho
 \end{align*}
where we used again that $\varphi$ is symmetric. Putting this all together gives
\begin{equation*}
    W*\rho = - M_0\rho -\frac{M_2 a^2}{2d}\Delta\rho +O(a^4M_4)
\end{equation*}
Using only the first two terms in the approximation in Eq. \eqref{eq:nonlocal_1species} yields
\begin{equation}
      \frac{\partial\rho}{\partial t} = -\nabla\cdot\left(\rho\nabla \left(\tilde{M}\Delta \rho + (M_0-\epsilon)\rho \right)\right),
      \label{approximatedPDE}
\end{equation}
with $\tilde{M} = M_2 a^2/2d$. Note that the approximation makes sense as long as moments $M_n$ of higher order ($n\geq4$) are small compared to $M_2$, and the scaling parameter $a$ is small. We emphasize here though, that the goal of our paper is not to compare \eqref{approximatedPDE} with the nonlocal model \eqref{eq:nonlocal_1species}, but to study possible behaviours of the local model in Eq. \eqref{approximatedPDE}.

Two relevant observations can be made here. First, note that the sign of $M_0-\epsilon$ gives the relative strength of repulsive and attractive forces. For negative $M_0-\epsilon$, localized repulsion is the dominant interaction, while for positive values of $M_0 - \epsilon$, the short-range attractive forces overcome repulsion. Here, we will focus on the latter case, since it is the biologically interesting one. In fact, with our choice of diffusion and aggregation potential $W$, Eq. \eqref{eq:nonlocal_1species} only has stationary states in the $M_0-\epsilon > 0$ case \cite{BurgerFranekSStates}. As we will see, our analysis here suggests that this is also the case for the local model given by Eq. \eqref{approximatedPDE}.

Secondly, and as it was already remarked in \cite{ArmstrongPainterSherratt}, the fourth order term has a dampening effect on the PDE. In the absence of this term, one obtains an ill-posed problem due to the negative diffusion coefficient. Therefore, it does not seem possible to have a second-order model of cell-cell adhesion, thus making evident the need for a fourth-order approximation. A similar phenomenon happens in \cite{negativeDiffusionAnguige} when one takes the continuum limit of a microscopic model incorporating cell-cell adhesion.

Before moving onto further considerations, and in order to facilitate the analysis, we nondimensionalize Eq. \eqref{approximatedPDE}. Under a suitable rescaling -- for instance, set $\rho\mapsto \tilde{M}\rho$ and $\mu^2 = (M_0-\epsilon)/\tilde{M}$ -- the model can be written as
\begin{equation}
    \frac{\partial\rho}{\partial t} = -\nabla\cdot\left(\rho\nabla \left(\Delta \rho + \mu^2\rho \right)\right),
    \label{approximatedPDE_rescaled}
\end{equation}
with $\mu^2 > 0$, according to our previous considerations. This model resembles a Cahn-Hilliard \cite{elliott1996cahn} or thin-film type equation where the parameter $\mu^{-2}$ plays the role of surface tension \cite{myers1998thin}, which somehow brings up again the idea of describing tissues using fluid-like properties, as originally proposed by Steinberg in his DAH. These considerations will become more relevant later on when we discuss systems of two species.

The thin-film equation describes the evolution of the thickness of a thin fluid that is lying on a surface. Equations of the type of Eq. \eqref{approximatedPDE_rescaled} appear as the lubrication approximation of a gravity-driven Hele-Shaw cell \cite{HeleShaw2,goldstein1998instabilities}. Depending on the sign of $\mu^2$, the equation is referred as \emph{long-wave unstable} or \emph{long-wave stable}. The sign of $\mu^2$ characterizes the linear stability of the constant steady state, but more details will be discussed later. This model also falls under a larger family of thin-film equations, whose properties have been well-studied -- see \cite{linearstabilitypugh,LAUGESEN2002377,laugesenpugh} for an exhaustive study of the steady states, \cite{slepvcev2009linear} for stability of self-similar solutions, and  \cite{bertozzi2,bertozzi1998long} for long-time behaviour of solutions and regularity.

Associated with the local model, we also have the Cahn-Hilliard type free energy
\begin{equation}
    \mathcal{F}[\rho]= \frac{1}{2}\int_\Omega\left(|\nabla\rho|^2 -\mu^2\rho^2\right) \mathrm{d}\mathbf{x}.
    \label{energyCH}
\end{equation}
With this in mind, \eqref{approximatedPDE_rescaled} can be written as a gradient flow with respect to the 2-Wasserstein metric (see for instance \cite{carrillo2003kinetic,matthes2009family,santambrogio2015optimal})
\begin{equation*}
    \frac{\partial\rho}{\partial t} = \nabla\cdot\left(\rho\,\nabla \frac{\delta \mathcal{F}}{\delta\rho}\right).
\end{equation*}
Observe that by integrating by parts formally, it holds
\begin{align*}
    \frac{\mathrm{d}}{\mathrm{d}t}\mathcal{F}[\rho] &= \int_\Omega \nabla\rho\cdot \nabla\left(\frac{\partial\rho}{\partial t}\right)\mathrm{d}\mathbf{x} - \mu^2\int_\Omega\rho\,\frac{\partial\rho}{\partial t}\,\mathrm{d}\mathbf{x}\\&=\int_\Omega\rho\,\nabla\left(\Delta\rho + \mu^2\rho\right)\cdot\nabla\left(\frac{\delta\mathcal{F}}{\delta\rho}\right)\mathrm{d}\mathbf{x} \\&=-\int_\Omega\rho\left|\nabla\frac{\delta\mathcal{F}}{\delta\rho}\right|^2\mathrm{d}\mathbf{x}\leq 0\,,
\end{align*}
and hence the energy is non-decreasing in time. Note that here we used the first and third order boundary conditions
\begin{equation}
    \label{eq:boundary_conditions}
    \partial_\nu\rho = \partial_\nu\Delta \rho = 0\quad\mbox{on }\partial\Omega,
\end{equation}
where $\nu$ is the exterior normal of $\Omega$.

\subsection{Linear stability analysis}
\label{section:linear_stability_analyis}
Equation \eqref{approximatedPDE_rescaled} admits as steady states any spatially homogeneous solution. One of the first biologically relevant questions that arises from this model, is whether aggregations are possible as in the case of Eq. \eqref{eq:nonlocal_1species}. To investigate this question we perform linear stability analysis on the spatially homogeneous solution $\rho(\textbf{x},t) = \rho_0$. In order to do so, we consider a perturbation $\rho(\textbf{x},t) = \rho_0 + \Tilde{\rho}(\textbf{x},t)$ and linearize the resulting equation. By setting $\Tilde{\rho}(\textbf{x},t)\propto e^{i\textbf{k}\cdot \textbf{x}+\sigma(\textbf{k})t}$ one finds the dispersion relation
\begin{equation*}
    \sigma(\textbf{k}) = \rho_0|\textbf{k}|^2\left(\mu^2-|\textbf{k}|^2\right).
\end{equation*}
In fact, the resulting linearized equation is identical as in the case of the standard Cahn-Hilliard equation describing phase separation in binary mixtures \cite{elliottCH}.
A necessary condition for the formation of non-trivial stationary states is then $\text{Re}(\sigma(\textbf{k})) > 0$ for certain values of the wave vector $\textbf{k}$, which results in the upper bound: $|\textbf{k}| < \mu$.

 From the unstability condition we already see that in the case where $\mu^2<0$, the homogeneous steady state is linearly stable and thus aggregation is not possible. As mentioned earlier, this case happens when $-\int_{\mathbb{R}^d}W < \epsilon$, for which Eq. \eqref{eq:nonlocal_1species} has no stationary states either \cite{BurgerFranekSStates}. We remark here that these considerations are not new, since Eq. \eqref{approximatedPDE_rescaled} falls under a larger family of thin-film equations, whose linear stability is well-known \cite{laugesenpugh}. From now on, we always consider the local model in the long-wave unstable regime $\mu^2>0$.

\subsection{Numerical experiments}
In this section, we explore numerically some basic properties of the local model. For that purpose we use a numerical scheme based on that in \cite{bailo2021unconditional}, which we briefly outline in \cref{sec:numerical_scheme}. Moreover, we run all of our simulations on a domain $[-L,L]$, where $L$ is specified individually for every experiment. We also assume periodic boundary conditions on $\rho$ and its derivatives.

\begin{figure}
    \centering
    \includegraphics[width = \textwidth]{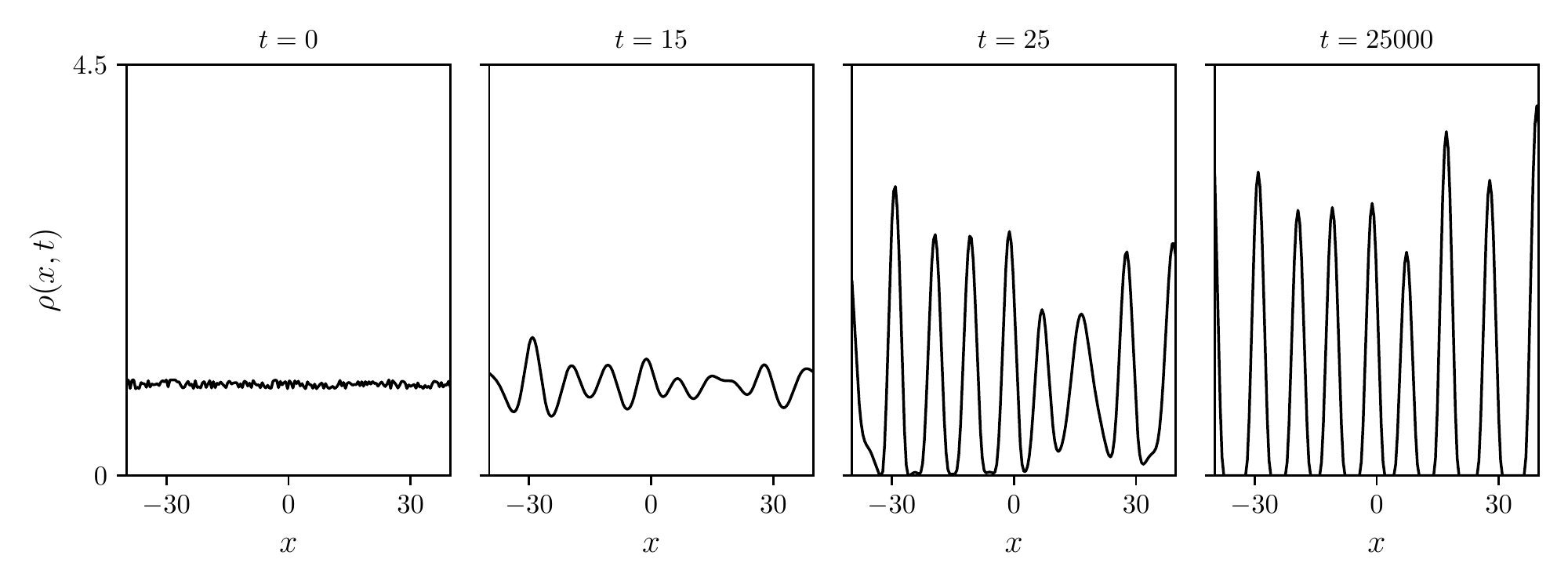}
    \caption{Aggregation is possible in the local model as long as $\mu^2>0$. Numerical simulations with periodic boundary conditions and parameters: $\mu^2 = 1, L = 40,\Delta x = 0.2, \Delta t = 0.01$. Initial data corresponds to the spatially homogeneous steady state $\rho(x,0) = 1$ for $x\in[-L,L]$ plus a small perturbation. }
    \label{Fig:Aggregations1D}
\end{figure}

\begin{figure}
    \centering
    \includegraphics[width = \textwidth]{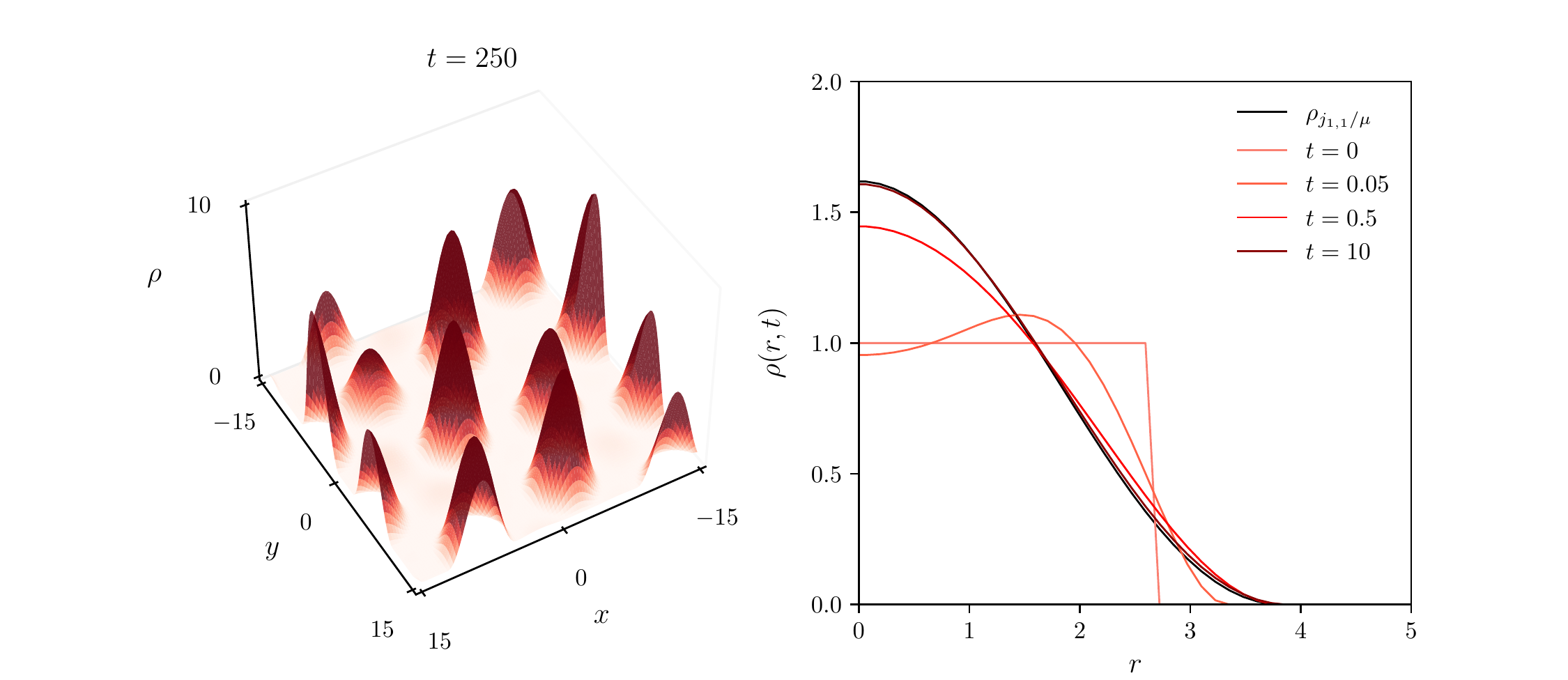}
    \caption{(left) Aggregation in the two-dimensional local model. Initial data is $\rho(x,0) = 1$ plus a small perturbation. Density configuration at $t = 250$. (right) Convergence to steady state in the two-dimensional model. Radial density profiles at different time points and stationary solution given by $\rho_{j_{1,1}/\mu}(r)$. Simulation parameters are $\Delta t = 0.01,\,\Delta x=0.1,\,\Delta y =0.1,\,\mu = 1$ and domain specifications: $L = 15$ for (a) and $L=5$ for $(b)$.}
    \label{Fig:Aggregations2D}
\end{figure}

We start by checking some of the results in the previous section and whether aggregations of cells are possible in this model. As expected from our derivations, the spatially homogeneous steady state is unstable in the case $\mu^2 > 0$. To test this prediction, we use a suitably large domain, $L = 40$, and perform simulations using as initial densities a slightly perturbed homogeneous steady state (Figure \ref{Fig:Aggregations1D}). We see that small perturbations rapidly lead to spatial patterning that mimic previous models of cell-cell adhesion \cite{ArmstrongPainterSherratt}. While cell densities change very rapidly at early times, as they evolve towards different peaks, smaller density bumps disappear at a very low rate, only reaching the stationary configuration after much longer times. In particular, the local model shows similar behaviour to nonlocal models with compactly supported interaction potentials, which usually give rise to stationary states with multiple separated aggregates \cite{nonlocalSchemceCarrilloYanghong}. The distance separating different cell aggregates is of course larger than the \emph{sensing radius} in the potential. A similar pattern appears in the two-dimensional case (Figure \ref{Fig:Aggregations2D}).

\subsubsection{Steady states and energy minimizers}

Next, we combine both analytical and numerical insights in order to study the stationary solutions of the local model.
Note that steady states of \eqref{approximatedPDE_rescaled} satisfy the equation
\begin{equation}
    \Delta \rho + \mu^2\rho = C,
    \label{steady_states1species}
\end{equation}
with $C$ a constant that could be different on each connected component of $\text{supp}(\rho)$. Observe that as a result of the gradient flow structure, calculating steady states is equivalent to the problem of finding critical points of the energy, Eq. \eqref{energyCH}, which are given by the condition: $-\delta\mathcal{F}/\delta\rho = C$.

Here, and motivated by our first numerical simulations, we assume that steady states are supported on finite unions of compact sets. In particular, and as we will see later in further numerical experiments, we assume that for smaller domains -- and not too disperse initial data -- there is only one connected component of $\text{supp}(\rho)$. This is what in the thin-film equation literature is called the \emph{droplet steady state} \cite{slepvcev2009linear}. We consider the one- and two-dimensional cases separately.

\textbf{One-dimensional case.} In the one-dimensional case, general solutions to Eq. \eqref{steady_states1species} read
\begin{equation*}
    \rho(x) = A\cos(\mu x)+B\sin(\mu x) + \frac{C}{\mu^2}.
\end{equation*}
Imposing that stationary states are both symmetric and invariant under translations, we can without loss of generality set $B = 0$. We now write $\text{supp}(\rho) = [-b,b]$ and use mass conservation to find
\begin{equation}
    \rho_b(x) = A\left(\cos(\mu x)-\cos(\mu b)\right), \quad A = \frac{m\mu}{2\left(\sin(\mu b)-\mu b\cos(\mu b)\right)},\label{steadystatesCH1D}
\end{equation}
where $m = \int_\Omega \rho(x,0)\mathrm{d}x$. With this, one finds a family of compactly supported steady states parametrized by $b$. Analogously, we could parametrize these steady states by their \emph{touchdown angle}, given by $\rho'(b)$. Note that in order to preserve positivity of solutions we need $\mu b\in(0,\pi]$.

However, numerical solutions show that for a given mass $m$, and in the cases where $\text{supp}(\rho)$ has only one connected component, solutions of  Eq. \eqref{approximatedPDE} tend to a unique steady state. We conjecture here that this steady state corresponds to the energy minimizer. In order to find it, we calculate the energy given by Eq. \eqref{energyCH} of the family of steady states in Eq. \eqref{steadystatesCH1D}
\begin{align*}
    \mathcal{F}[\rho_b]
    =\frac{m^2\mu^3}{2}\frac{\cos(\mu b)}{\sin(\mu b)-\mu b\cos(\mu b)}.
\end{align*}
 Note that $\mathcal{F}[\rho_b]$ is a decreasing function of $\mu b$ on $(0,\pi]$ and hence the energy minimizer corresponds to the case where $\mu b = \pi$:
 \begin{equation*}
     \rho_{\pi/\mu}(x) = \frac{m \mu}{2\pi}\left(\cos(\mu x) + 1\right), \quad |x| \leq \frac{\pi}{\mu}\,.
 \end{equation*}

 Observe here that the minimum of the energy $\mathcal{F}[\rho_b]$ is obtained when the zero contact angle condition $\rho'(b) = 0$ is satisfied, in the same way as in \cite{BernoffTopazCH}. We emphasize here that while we have been able to identify the steady state with the lowest energy, in general this is a very complex problem -- see \cite{LAUGESEN2002377} for an exploration of the energy landscape for a larger family of thin-film type equations.

\textbf{Two-dimensional case.}
The calculations here are very similar to the one dimensional case. Again based on numerical simulations we assume that steady states have radial symmetry and write $r = |\mathbf{x}|$. With this, general solutions $\rho = \rho(r)$ to Eq. \eqref{steady_states1species} read
\begin{equation*}
    \rho(r)=AJ_0(\mu r)+BY_0(\mu r) + \frac{C}{\mu^2}\,,
\end{equation*}with $J_n$ and $Y_n$ being Bessel functions of the first and second kind, respectively. Imposing regularity at the origin yields $B = 0$ and, once again, assuming that steady states are supported on a closed disk of radius $b$, we obtain
\begin{equation*}
    \rho_b(r) = A\left(J_0(\mu r) - J_0(\mu b)\right),\quad A = \frac{m\mu}{\pi b\left(J_1(\mu b)-\mu bJ_0(\mu b)\right)}\,.
\end{equation*}
For positive solutions we need to impose that $\mu b\in (0,j_{1,1}]$ where $j_{1,1}\approx 3.832$ is the first zero of $J_0' = -J_1$. The energy minimization argument also holds here. Note that
\begin{equation*}
    \mathcal{F}[\rho_b] = \frac{m^2\mu^4}{\pi}\frac{J_0(\mu b)}{\mu b\left(2J_1(\mu b)-\mu b J_0(\mu b)\right)}\,,
\end{equation*}
is a decreasing function of $\mu b$ on $(0,j_{1,1}]$, and hence the minimizer corresponds to the case $\mu b = j_{1,1}$ -- or equivalently to the solution satisfying $\rho'(b) = 0$. This steady state reads
\begin{equation*}
    \rho_{j_{1,1}/\mu} (r) = \frac{m\mu^2}{\pi j_{1,1}^2}\left(1-\frac{J_0(\mu r)}{J_0(j_{1,1})}\right), \quad |x| \leq \frac{j_{1,1}}{\mu}\, ,
\end{equation*}
with $J_0(j_{1,1})\approx -0.403$. In Figure \ref{Fig:Aggregations2D} we sketch the radial density profiles for simulations in a two-dimensional box of length $2L = 10$.

\subsection{Contact angle of energy minimizers}
Here we have found that the energy minimizer satisfies the zero contact angle condition $\rho'(b) = 0$. In fact, this can be justified using a perturbation argument in one dimension as in \cite{BernoffTopazCH}, and without the need for explicit expressions for the steady states. To see this, assume that $\rho$ is a symmetric solution with compact support given by $\text{supp}(\rho) = [-b,b]$. Now perturb the support $\Bar{b} = b + \delta b$, with $\delta b\ll b$ and assume that the solution is also perturbed according to $\Bar{\rho} =\rho + \delta\rho$. Note that the condition $\int \Bar{\rho} = \int \rho$ requires the perturbation to have zero total mass. If we calculate the free energy of the new solution we obtain up to first order in the perturbation
\begin{equation*}
    \mathcal{F}[\Bar{\rho}] = \mathcal{F}[\rho] + \int_{-b}^{b}\delta\rho\,\frac{\delta\mathcal{F}}{\delta\rho} \,\mathrm{d} x + \delta b\cdot\rho'(b)^2.
\end{equation*}
If we assume that $\rho$ is a minimizer, then the second term becomes zero, as $\rho$ satisfies Eq. \eqref{steady_states1species}. If $\rho'(b)\neq 0$ then we can find solutions with lower energy, contradicting the fact that $\rho$ is a minimizer. Hence $\rho'(b) = 0$.

\section{Extension to two interacting populations}

Having analyzed the model for a single population, we now extend it to two interacting cell populations. Our main goal here is to study if such model is able to reproduce the patterns seen in the Steinberg experiments, and whether this behaviour can be understood in terms of the model parameters.

\begin{figure}
    \centering
    \includegraphics[width = .95\textwidth]{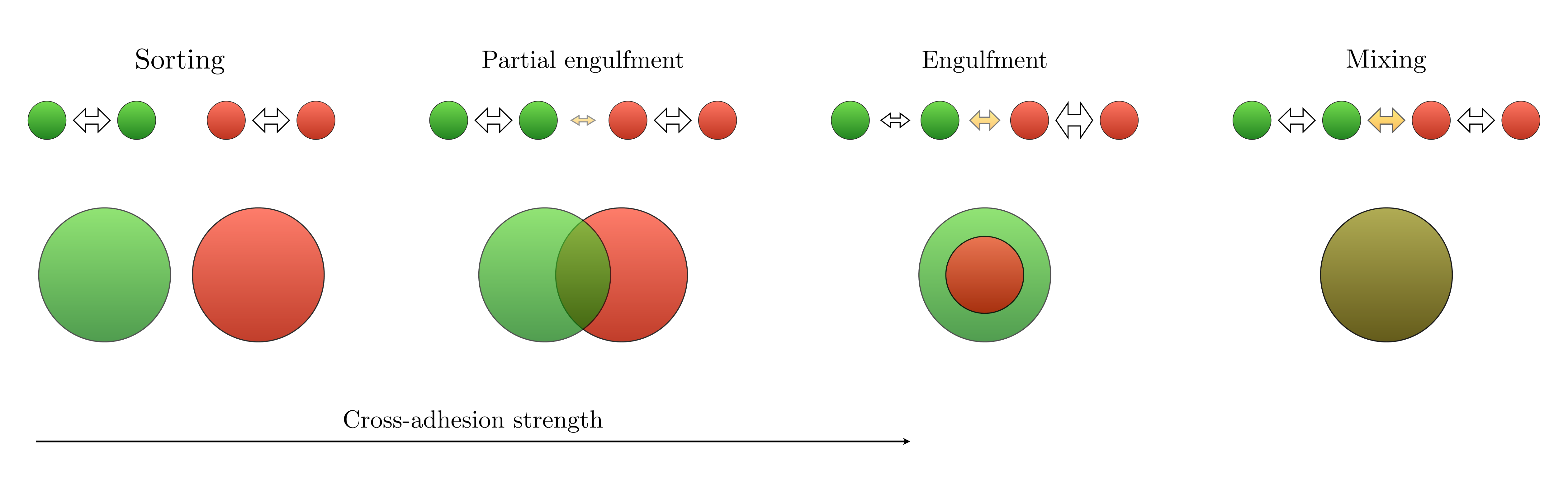}
    \caption{Possible configurations for Steinberg experiments in terms of the cross-adhesion and the self-adhesion of a system of two species (adapted from \cite{MurakawaTogashi}). In the weak cross-adhesion regime we might have two patterns depending on whether the cross-adhesion strength is strictly zero or positive. \textbf{Sorting} is observed when there is no cross-adhesion between the two species, and \textbf{partial engulfment} when cross-adhesion is small compared to the self-adhesion of each population. When the cross-adhesion is stronger, the system might evolve to an \textbf{engulfment} pattern, where the more cohesive species is surrounded by the less cohesive one; or to complete \textbf{mixing} of the cell populations. The first corresponds to the case in which the cross-adhesion is stronger than the self-adhesion of one species but weaker than the self-adhesion of the other one. The latter occurs when the cross-adhesion strength is comparable to both self-adhesion forces.}
    \label{fig:DAH_twospecies}
\end{figure}

In the case of two interacting populations, the self-adhesion of each species and the cross-adhesion between them determine the behaviour of the system. Depending on the relative strength of adhesive forces, experimentally it is seen that the two cell populations may evolve to one of four different configurations, that we represent in Figure \ref{fig:DAH_twospecies}.

\subsection{A system of thin-film equations to model cell-cell adhesion}

We proceed as in the one species case and assume that the potentials $W_{ij}$  in Eqs. \eqref{eq:nonlocal_2species}  are attractive and scale according to a parameter $a$, which gives the range of interactions. More precisely we impose $W_{ij}(\textbf{x})=-a^{-d}\varphi_{ij}\left(\textbf{x}/a\right)$, with the functions $\varphi_{ij}$ satisfying the conditions in the previous section. In the limit $a\rightarrow 0$, we can approximate $W_{ij}*f\approx - c_{ij}f -d_{ij}\Delta f$, where $f \in\{ \rho,\eta\}$ and the constants $c_{ij},d_{ij}$ could be different for each potential. Note that $c_{ij}$ is the volume of $\varphi_{ij}$ and $d_{ij}$ is related to its second moment \begin{equation*}
    c_{ij} = \int_{\mathbb{R}^d} \varphi_{ij}(\textbf{x})\,\mathrm{d}\textbf{x},\quad d_{ij} = \frac{a^2}{2d}\int_{\mathbb{R}^d} |\textbf{x}|^2\varphi_{ij}(\textbf{x})\,\mathrm{d}\textbf{x}.
\end{equation*} For simplicity we assume here that the cross-interaction potential is the same for the two species $W_{12}=W_{21}$, which is a commonly used assumption in many models of cell-cell adhesion \cite{CarrilloMurakawaCellAdhesion, ArmstrongPainterSherratt}. Using these approximations in the two species nonlocal model Eqs. \eqref{eq:nonlocal_2species}, yields

\begin{subequations}
\begin{align*}
     \frac{\partial \rho}{\partial t} = -\nabla\cdot\left(\rho\nabla \left(\kappa_1\Delta \rho + \tilde{\alpha}\Delta \eta + \mu_1\rho + \tilde{\omega}\eta \right)\right),
     \\
     \frac{\partial \eta}{\partial t} = -\nabla\cdot\left(\eta\nabla \left(\tilde{\alpha}\Delta \rho + \kappa_2\Delta \eta + \tilde{\omega}\rho + \mu_2\eta \right)\right).
\end{align*}
\end{subequations}

The model parameters can be understood in terms of the potentials $W_{ij}$. First note that the parameters in the fourth order terms, $\kappa_1,\kappa_2$ and $\tilde{\alpha}$ are directly related to the second moments of the potentials. Hence, they only give information on the strength and range of attractive forces. Assuming that the potentials are all attractive, we have $\kappa_1,\kappa_2,\alpha\geq 0$. On the other hand, the parameters in the second order terms, $\mu_1,\mu_2$ and $\tilde{\omega}$, are both related to the volumes of the potentials, and to the strength of repulsive forces, which are given by $\epsilon$. According to our considerations in Section \ref{section:linear_stability_analyis}, we assume that $-\int_{\mathbb{R}^d}W_{11} > \epsilon$ and also $-\int_{\mathbb{R}^d}W_{22} > \epsilon$, meaning that self-attraction overcomes repulsion in each  of the  populations. This gives $\mu_1,\mu_2>0$. However, cross-attraction between the two type of cell types could be weaker and thus $\tilde{\omega}$ could be either positive or negative.

In fact, and in order to facilitate the analysis, we can reduce the number of parameters with a suitable rescaling of the variables. For example, set $x \mapsto \xi x$, $t \mapsto T t$, $\rho\mapsto\mu_2\rho$, $\eta\mapsto\mu_2\eta$ with $\xi^2 = {\mu_2}/{\kappa_2}$ and $T = \xi^{2}$. This yields the new rescaled parameters
\begin{equation*}
     \kappa = \frac{\kappa_1}{\kappa_2}\,,\quad\alpha = \frac{\tilde{\alpha}}{\kappa_2}\,,\quad\mu = \frac{\mu_1}{\mu_2}\,,\quad\omega = \frac{\tilde{\omega}}{\mu_2}\,,
\end{equation*}
and the reduced system
\begin{subequations}
\begin{align}
     \frac{\partial \rho}{\partial t}& = -\nabla\cdot\left(\rho\nabla \left(\kappa\Delta \rho + \alpha\Delta \eta + \mu\rho + \omega\eta \right)\right),\label{eq:local_2species_a}\\
     \frac{\partial \eta}{\partial t}& = -\nabla\cdot\left(\eta\nabla \left(\alpha\Delta \rho + \Delta \eta + \omega\rho + \eta \right)\right)\label{eq:local_2species_b}.
\end{align}\label{eq:local_2species}
\end{subequations}
Here, $\kappa\geq 0$ and $\mu> 0$ represent the relative self-adhesion strength of $\rho$ with respect to $\eta$; while $\alpha\geq 0$ and $\omega\in\mathbb{R}$ give the relative strength of the cross-attraction forces.

Observe too that the local model, Eqs. \eqref{eq:local_2species}, is essentially a system of two thin-film like equations, where the parameters $\kappa$ and $\alpha$ can be understood as the relative tensions, of one species with respect to the other one, and of the interface separating the two populations. The parameters in the second order terms $\mu$ and $\omega$ are then related to the population pressure exerted by each cell type. Under this setting, it makes sense to ask whether differential tension as thought originally by Steinberg is able to explain cell sorting phenomena. In other words, can we identify relevant regimes for the four parameters $\kappa,\alpha,\mu,\omega$ such that the experimental patterns are recovered in the local model? Note here that the local model given in Eqs. \eqref{eq:local_2species} is not a phenomenological description emerging from the tissue-fluid analogy, but arises in the limit of short-range interactions of Eqs. \eqref{eq:nonlocal_2species}, and hence it provides a direct connection between the original DAH and a model of cell-cell adhesion derived from first physical principles.

In order to avoid negative diffusion, we require the matrix
\begin{equation*}
    M = \begin{pmatrix}
 \kappa & \alpha\\
 \alpha & 1\\
\end{pmatrix},\label{eq:M}
\end{equation*}
to be positive definite. Since $\kappa > 0$, this requires $\det M \geq 0$. This sets a limit on the strength of the cross-attraction $0\leq \alpha<\sqrt{\kappa}$.
\subsection{Energy}

Thanks to the symmetry in the cross-interaction terms given by $\alpha$ and $\omega$, this system also exhibits a gradient-flow structure
\begin{subequations}
\begin{align*}
     \frac{\partial\rho}{\partial t} = \nabla\cdot\left(\rho\,\nabla \frac{\delta \mathcal{F}_2}{\delta\rho}\right),\\
         \frac{\partial\eta}{\partial t} = \nabla\cdot\left(\eta\,\nabla \frac{\delta \mathcal{F}_2}{\delta\eta}\right),
\end{align*}
\end{subequations}
with respect to the  2-Wasserstein metric \cite{carrillo2003kinetic,matthes2009family,santambrogio2015optimal} and the Cahn-Hilliard type free energy
\begin{equation}
    \mathcal{F}_2[\rho,\eta] = \int_\Omega \left(\frac{\kappa}{2}|\nabla\rho|^2+\frac{1}{2}|\nabla\eta|^2+\alpha\nabla\rho\cdot\nabla\eta -\frac{\mu}{2}\rho^2-\frac{1}{2}\eta^2-\omega\rho\eta\right) \mathrm{d}\mathbf{x}.\label{eq:energy2}
\end{equation}
We remark here that the nonlocal model for two species given by Eqs. \eqref{eq:nonlocal_2species} also exhibits a gradient flow structure when the cross-interaction potentials are symmetrizable, which provides with variational schemes to prove the existence of solutions of the system \cite{difrancesco2013measure,AntonioTwoSpeciesNonlocal}.

As in the one-species case, we can formally integrate by parts to show that the energy is non-increasing in time
\begin{align*}
     \frac{d}{dt}\mathcal{F}_2[\rho,\eta]  =&\,\kappa\int_\Omega \nabla\rho\cdot \nabla\left(\frac{\partial\rho}{\partial t}\right)\mathrm{d}\mathbf{x} +   \int_\Omega \nabla\eta\cdot \nabla\left(\frac{\partial\eta}{\partial t}\right)\mathrm{d}\mathbf{x}
     \\
     &+ \alpha\int_\Omega\nabla\rho\cdot\nabla\left(\frac{\partial\eta}{\partial t}\right)\mathrm{d}\mathbf{x}
     +\alpha\int_\Omega\nabla\left(\frac{\partial\rho}{\partial t}\right)\cdot\nabla\eta \,\mathrm{d}\mathbf{x}
     \\
     &-\mu\int_\Omega\rho\,\frac{\partial\rho}{\partial t} \,\mathrm{d}\mathbf{x} -\int_\Omega\eta\,\frac{\partial\eta}{\partial t}\, \mathrm{d}\mathbf{x}
     -\omega\int_\Omega\rho\,\frac{\partial\eta}{\partial t} \,\mathrm{d}\mathbf{x} -\omega\int_\Omega\frac{\partial\rho}{\partial t}\,\eta \,\mathrm{d}\mathbf{x}
     \\
     =&\,\int_\Omega\rho\,\nabla\left(\kappa\Delta\rho +\alpha\Delta\eta+\mu\rho+\omega\eta\right)\cdot\nabla\left(\frac{\delta\mathcal{F}_2}{\delta\rho}\right)\mathrm{d}\mathbf{x}
     \\
     &+\int_\Omega\eta\,\nabla\left(\alpha\Delta\rho +\Delta\eta+\omega\rho+\eta\right)\cdot\nabla\left(\frac{\delta\mathcal{F}_2}{\delta\eta}\right)\mathrm{d}\mathbf{x}
     \\
     =&\,-\int_\Omega\rho\left|\nabla\frac{\delta\mathcal{F}_2}{\delta\rho}\right|^2\mathrm{d}\mathbf{x}-\int_\Omega\eta\left|\nabla\frac{\delta\mathcal{F}_2}{\delta\eta}\right|^2\mathrm{d}\mathbf{x}\leq 0.
\end{align*}
Again, we used the boundary conditions on $\rho$ and $\eta$ given by Eq.  \eqref{eq:boundary_conditions}.

\subsection{Numerical simulations for Steinberg experiments in one dimension}

Here we study numerically whether the local model is able to reproduce the four different patterns observed in Steinberg experiments (Figure \ref{fig:DAH_twospecies}), namely: (i) mixing; (ii) engulfment; (iii) partial engulfment; (iv) sorting. Animated movies of the simulations in this section are available in \cite{figshare}. To do that, one must understand what parameter ranges should correspond to each one of the observed patterns. This is simpler when one assumes a particular shape for the potentials $\varphi_{ij}$. Let us then assume that these only differ by constants, i.e. $\varphi_{ij} = K_{ij}\varphi$, for constants $K_{ij}\geq 0$ satisfying $K_{12} = K_{21}$ and a given potential $\varphi$. This is actually the case in previously used nonlocal models \cite{ArmstrongPainterSherratt,CarrilloMurakawaCellAdhesion,carrillo2018zoology}, where the constants $K_{ij}$ give the adhesive strengths of the two cell populations. Under these assumptions, the model parameters are directly related to the moments of $\varphi$ and the constants $\epsilon$ and $K_{ij}$:
\begin{equation*}
    \kappa = \frac{K_{11}}{K_{22}}\,,\quad\alpha = \frac{K_{12}}{K_{22}}\,,\quad\mu =\left( \frac{M_0-\epsilon/K_{11}}{M_0-\epsilon/K_{22}}\right)\kappa,\quad\omega =\left( \frac{M_0-\epsilon/K_{12}}{M_0-\epsilon/K_{22}}\right)\alpha,
   \end{equation*}
   where $M_0$ is the volume of the potential $\varphi$.
    Note then that $\kappa$ and $\alpha$ can be interpreted as relative adhesion strengths as mentioned earlier. However, $\mu$ and $\omega$ are not only related to cell-cell adhesion but also to the strength of local repulsion due to volume exclusion.

    \begin{figure}
     \centering
     \includegraphics[width = \textwidth]{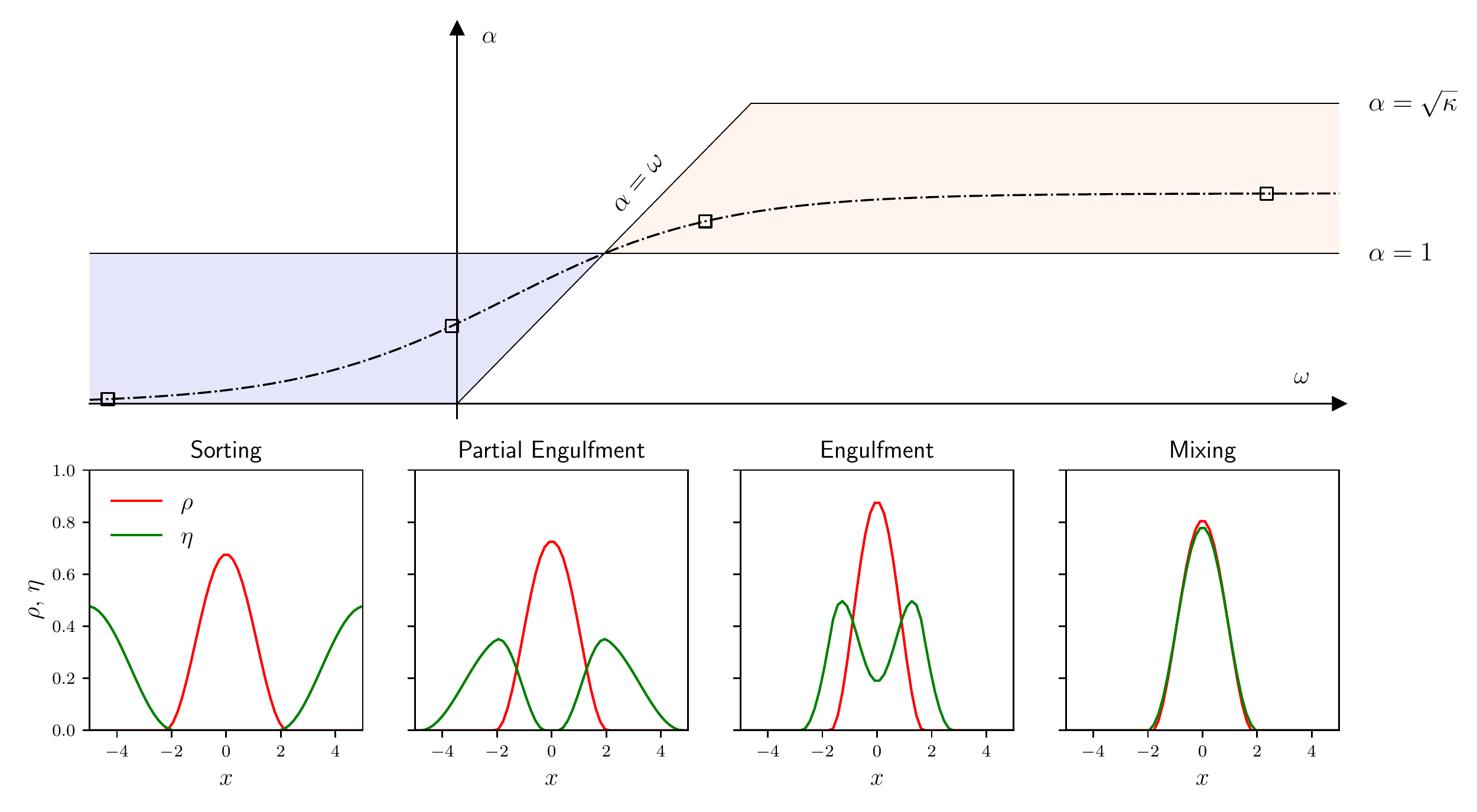}
     \caption{Understanding the impact of changing model parameters. Imposing that $\eta$ is the less cohesive population implies $\mu>\kappa>1$ as discussed in the text. We focus then  on the cross-interactions. Parameter ranges for $\alpha$ and $\omega$ shown above: the blue-shaded region represents the weak cross-adhesion regime, while the red-shaded region corresponds to the case of strong cross-adhesion. Below we plot the numerically found steady states for the parameter values given by the square points: sorting, $\omega = -2.38,\,\alpha =0.03$; partial engulfment, $\omega = -0.04,\,\alpha =0.52$; engulfment, $\omega = 1.69,\,\alpha = 1.21$; mixing, $\omega = 5.51,\,\alpha =1.40$. In every case $\kappa = 2$ and $\mu = 4$. We observe the different patterns seen in the Steinberg experiments and the transition from sorting to mixing as we increase the cross-adhesion, agreeing with the model interpretation. Numerical simulations performed on a domain of length $L = 5$ and $\Delta x = 0.2,\,\Delta t = 0.01$ with periodic boundary conditions and initial condition $\rho(x,0) = \eta(x,0) = \chi_{|x|<1.5}/2$. See \cite{figshare} for an animated movie with the stationary states corresponding to each point in the dashed line.}
     \label{fig:4patterns}
 \end{figure}

    We also assume without loss of generality that $\eta$ is the less cohesive population and hence $K_{22}<K_{11}$. Then according to these expressions we have $\mu>\kappa>1$. Parameter values outside of this range are also valid but  interpreting the model in such cases becomes more challenging. For the cross-interaction parameters $\alpha$ and $\omega$, one must distinguish two regimes depending on $K_{12}$, i.e. the strength of the cross-adhesion:
    \begin{enumerate}
        \item \text{Weak cross-adhesion} ($K_{12}<K_{22})$. In this case, we have $\omega < \alpha < 1$, which is given by the blue-shaded region in Figure \ref{fig:4patterns}. The smaller the cross-adhesion strength, the more negative $\omega$ becomes and the smaller $\alpha$ is too.

        \item \text{Strong cross-adhesion} ($K_{12}>K_{22})$. On the other hand, when the cross-adhesion is stronger than the self-adhesion of the second cell type, we have $\omega > \alpha > 1$. This is depicted in the red-shaded region in Figure \ref{fig:4patterns}. Now, the larger the cross-adhesion strength, the larger $\alpha$ and $\omega$ are. Note too that the quotient $\omega/\alpha$ is increasing with $K_{12}$, meaning that in the limit of strong cross-adhesion we should expect $\omega\gg\alpha$.
    \end{enumerate}

    Given the gradient flow structure of the local model in Eqs. \eqref{eq:local_2species}, one could also understand these regimes by looking at the free energy $\mathcal{F}_2[\rho,\eta]$. We focus on the interaction terms in Eq. \eqref{eq:energy2}. In particular, it becomes evident that whenever $\omega<0$, then in order to minimize the energy, both species will tend to separate so that the value of the product $\rho\eta$ is small. On the other hand, when $\omega>0$, the two cell types will be attracted to each other, trying to maximize the value of $\rho\eta$.

  We now simulate Eqs. \eqref{eq:local_2species} having in mind the above considerations. We start with numerical simulations in small domains and periodic boundary conditions, as shown in Figure \ref{fig:4patterns}. The figure suggests that our intuition of the model was correct, since the described regimes are able to replicate the four patterns observed in the Steinberg experiments. As predicted by the DAH, we observe that the two cell populations tend to separate when the cross-adhesion is weak. On the other hand, if the cross-adhesion is larger, the more cohesive population $\rho$ gets engulfed inside $\eta$, and eventually the cross-adhesion is strong enough to drive mixing of the two.

  \begin{figure}
    \centering
    \includegraphics[width = \textwidth]{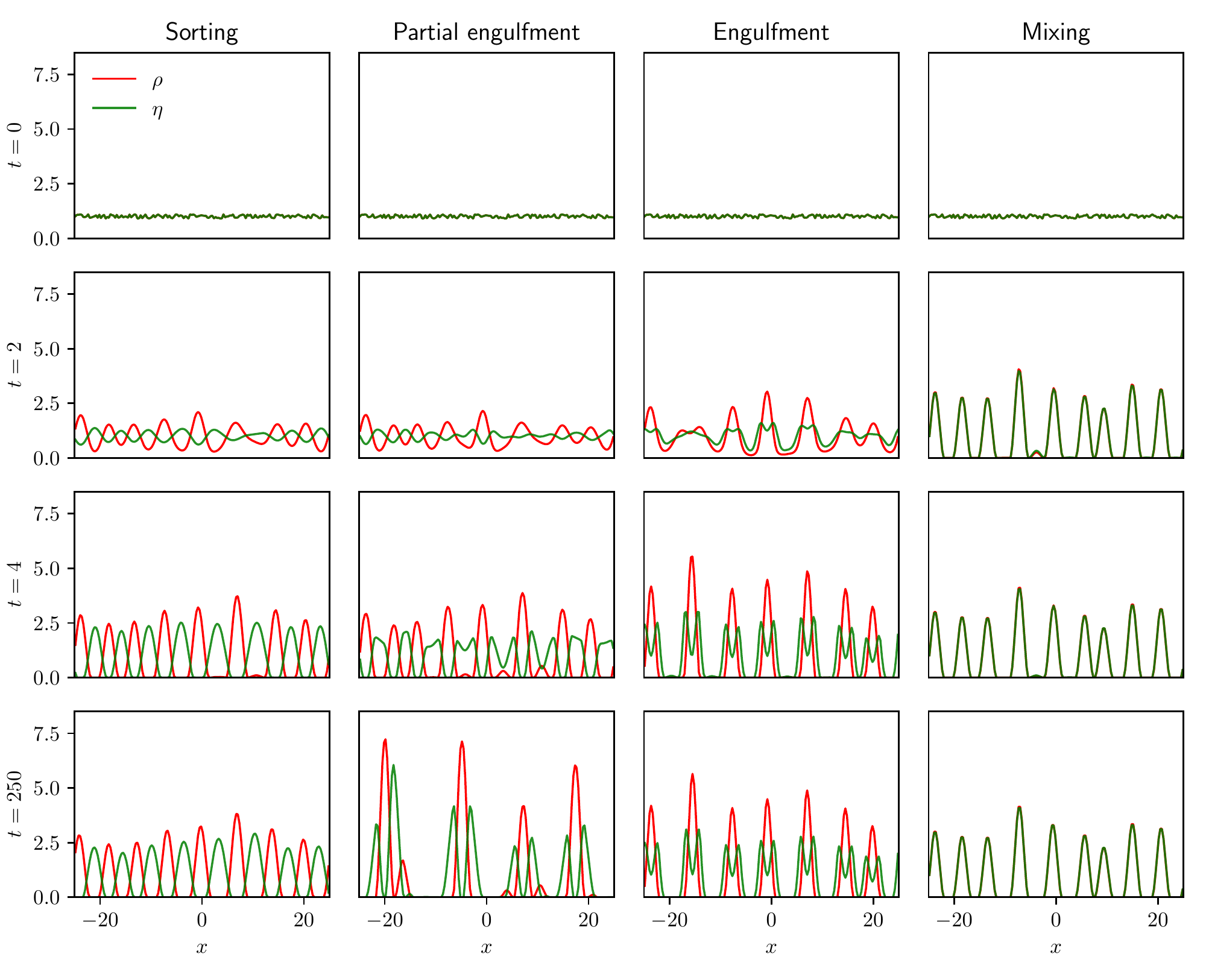}
    \caption{Solutions of the local model using model parameters related to the Steinberg experiments. Each column represents the solution with the same set of parameters and at different times. Mixing, $\alpha = 1.4 ,\,\omega =6$; engulfment, $\alpha = 1.3 ,\,\omega =2$; partial engulfment, $\alpha = 0.8 ,\,\omega = 0.2$; sorting, $\alpha = 0 ,\,\omega =-1$. In every case $\kappa = 2$ and $\mu = 4$ and also  $L = 25,\,\Delta x = 0.2,\, \Delta t =0.01$. See \cite{figshare} for animated movies.}
    \label{Fig:2species_1d}
\end{figure}

  The same patterns emerge in numerical simulations on larger domains. Here we choose parameter values corresponding to the two regimes that we described above, and show the solutions at different times in Figure \ref{Fig:2species_1d}. When the cross-adhesion is non-zero ($\alpha\neq 0$), steady states are composed of multiple compactly supported blobs or bumps. In the next section we will see that it is possible to find analytically the exact shape of each one of these bumps, given their individual masses. This is an advantage with respect to nonlocal models, where analytical solutions are only available for specific types of  potentials \cite{carrillo2018zoology}. Note however, that predicting the final mass of each of the bumps is difficult.

\begin{figure}
     \centering
     \includegraphics[width = .8\textwidth]{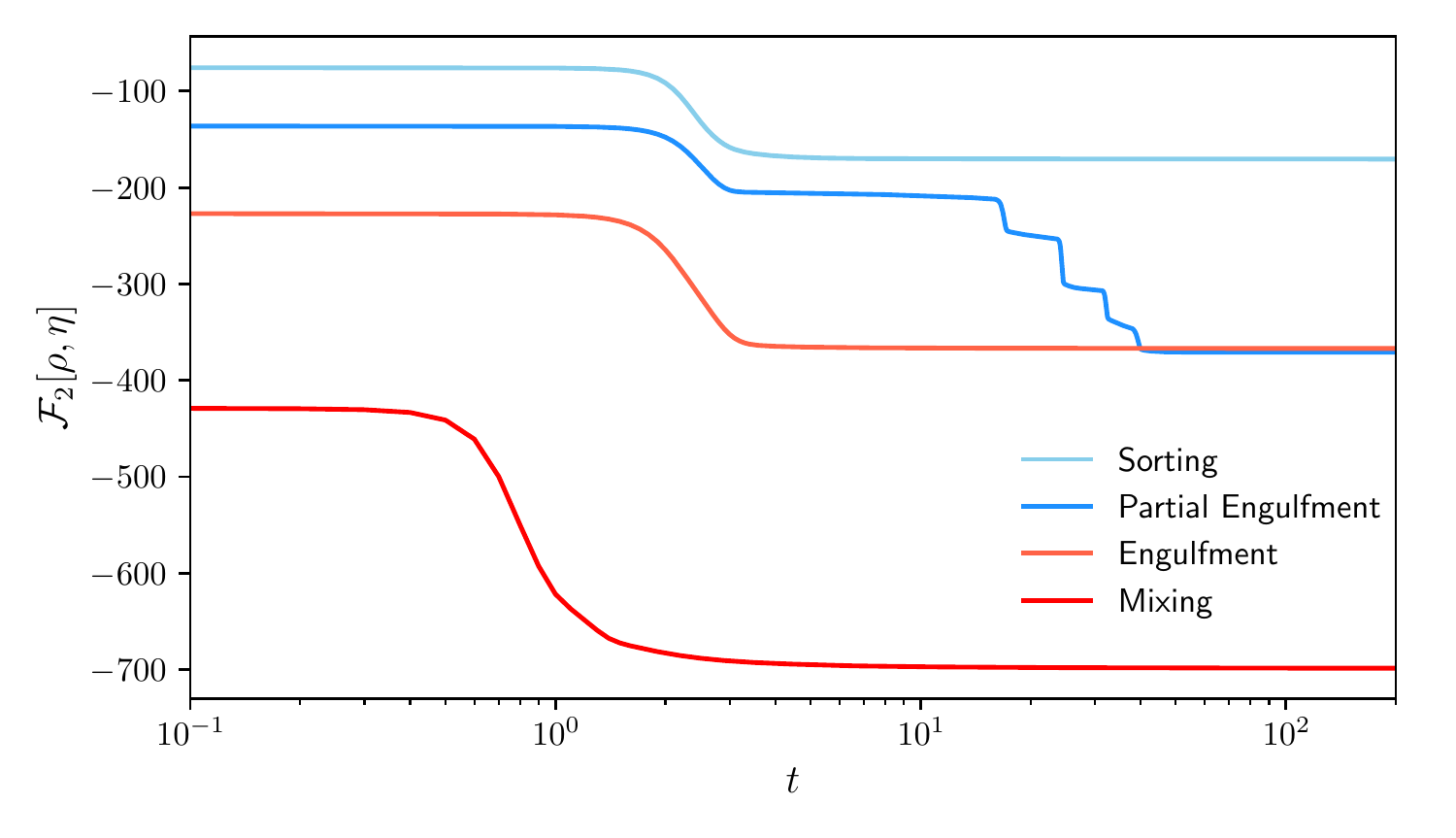}
     \caption{Energy decay given by Eq. \eqref{eq:energy2} for the numerical solutions in Figure \ref{Fig:2species_1d}. Solutions corresponding to the weak and strong cross-adhesion regimes represented in blue and red, respectively. In general, the stronger the cross-adhesion, the faster the decay of $\mathcal{F}_2[\rho,\eta]$.}
     \label{fig:energy}
 \end{figure}

Observe too that solutions in the weak cross-adhesion regime -- corresponding to the sorting and partial engulfment patterns -- show very similar behaviour for early times (see Figures \ref{Fig:2species_1d} and \ref{fig:energy}). When $\alpha = 0$ then both cell species tend to separate, converging to more or less sharply segregated solutions -- which does not happen in previous nonlocal models that consider linear diffusion \cite{ArmstrongPainterSherratt}. However, when cross-adhesion is small but strictly positive, the two populations move away from each other at early times and later organize themselves to form different aggregates, composed by different coexistence regions. This kind of metastability (see Figure \ref{fig:energy}) has also been observed before in Cahn-Hilliard type systems \cite{barrett2001fully,celora2021dynamics}.

 \subsection{Stationary solutions} We now move our attention to the study of stationary solutions of the local system, Eqs \eqref{eq:local_2species}. The computations in this section were performed with the help of the open-source software SageMath \cite{sagemath}.

 Motivated by the numerical simulations in the previous sections, we assume that such stationary solutions are compactly supported, and focus on one of the compact intervals shown in Figure \ref{Fig:2species_1d} where we have nonzero solutions. In particular we assume that stationary solutions are of the form of the solutions corresponding to the engulfment/mixing patterns shown in Figure \ref{fig:4patterns}. Let us assume again that $\rho$ is the more cohesive populations and thus, as discussed before, $\mu>\kappa$. Having this in mind, we consider
 $$\text{supp}(\rho) = [-b,b],\qquad \text{supp}(\eta) = [-c,c],$$
 with $0<b<c$. Note that this defines a coexistence region for the two species which is given by the support of $\rho$. For simplicity, we further assume that both species have the same mass $\int_{-b}^b\rho\,\mathrm{d}x = \int_{-c}^c\eta\,\mathrm{d}x = m$. For different masses, a similar calculation shows that the stationary solutions depend on each individual mass, although we expect the same biological phenomena to be possible across parameter space. A similar behaviour is also observed for nonlocal models \cite{carrillo2018zoology}.

 Stationary solutions of Eqs. \eqref{eq:local_2species} are characterized by
\begin{subequations}
\begin{align}
      \kappa \rho''& + \alpha\eta'' + \mu\rho+\omega\eta = C_1;\label{eq:steady1} \\
           \alpha \rho''& + \eta'' + \omega\rho+\eta = C_2;\label{eq:steady2}
\end{align}\label{eq:steady}
\end{subequations}
where $C_1$ and $C_2$ are arbitrary constants to be determined. Note again that these two conditions mean that steady states of Eqs. \eqref{eq:local_2species} are critical points of the energy $\mathcal{F}_2$. Here, and motivated by our exploration of the one species system, we conjecture that these steady states correspond to the energy minimizers, which in the two species case also satisfy the zero contact angle condition $\rho'(b) = \eta'(c) = 0$.

To see that, we follow the same argument, again motivated by \cite{BernoffTopazCH}, and consider $\Bar{b} = b+\delta b$, $\Bar{c} = c+\delta c$, $\Bar{\rho} = \rho+\delta \rho$, and $\Bar{\eta} = \eta+\delta \eta$. Calculating the perturbed energy up to first order we find
\begin{align*}
    \mathcal{F}_2[\Bar{\rho},\Bar{\eta}] =& \,\mathcal{F}_2[\rho,\eta] + \int_{-b}^{b}\delta\rho\,\frac{\delta\mathcal{F}_2}{\delta\rho}\, \mathrm{d} x +\int_{-c}^{c}\delta\eta\,\frac{\delta\mathcal{F}_2}{\delta\eta}\, \mathrm{d}x \\&+\delta b\cdot\kappa\rho'(b)^2+\delta c\cdot\eta'(c)^2+\delta b\cdot2\alpha\rho'(b)\eta'(b).
\end{align*}
Then, if $(\rho,\eta)$ minimizes the energy we need $\rho'(b) = \eta'(c) = 0$.

\subsubsection{Outside the coexistence region}

We thus look for symmetric solutions with zero contact angle. Using both assumptions and integrating Eqs. \eqref{eq:steady} on $[-b,b]$ we find
\begin{subequations}
\begin{align}
2b\,C_1 &= 2\alpha\eta'(b) +m\left(\mu+\omega\delta\right),\label{eq:C_1}\\2b\,C_2 &= 2\eta'(b) +m\left(\omega+\delta\right),\label{eq:C_2a}\end{align}\label{eq:C_1_C_2}
\end{subequations}
where $\delta = \int_{-b}^{b}\eta\,\mathrm{d}x/\int_{-c}^{c}\eta\,\mathrm{d}x$ is the fraction of mass of $\eta$ in the coexistence region. Integrating the equation for $\eta$, Eq. \eqref{eq:steady2}, on $[-c,c]$ we obtain
\begin{equation}
    2c \,C_2 =m\left(\omega+1\right) \label{eq:C_2b}.
\end{equation}
This last expression gives $C_2$ in terms of the model parameters and $c$. Note too that $\eta'(b)$ can be solved from Eqs.  \eqref{eq:C_1_C_2} and using Eq. \eqref{eq:C_2b}, we can also find $C_1$ in terms of the model parameters, $b$, $c$ and the mass fraction $\delta$.

We solve first for $\eta$, outside the coexistence region. When $x\in [b,c]$ we have $\rho = 0$ and then
$$\eta''+\eta = C_2.$$
General solutions read
\begin{equation}
    \eta(x) = A_1\cos(x)+B_1\sin(x) + C_2,
\end{equation} with $A_1, B_1$ constants to be determined. Solutions on $[-c,-b]$ can be found via the substitution $B_1\mapsto -B_1$, due to the symmetry assumption. Imposing $\eta(c) = \eta'(c) = 0$ and using \eqref{eq:C_2b} we find an explicit expression for $\eta$
\begin{equation*}
    \eta(x) = \frac{m\left(\omega+1\right)}{2c}\left(1-\cos\left(c-\frac{x^2}{|x|}\right)\right),\qquad\text{for } |x|\in[b,c].
\end{equation*}
Note that $c$ is still unknown. However, knowing $\eta$ outside the coexistence region is enough to find also the mass fraction $\delta$ in terms of $b$ and $c$ and model parameters
\begin{equation*}
    \delta = \frac{\int_{-b}^b\eta\,\mathrm{d}x}{\int_{-c}^c\eta\,\mathrm{d}x} = 1-\frac{2}{m}\int_b^c\eta\,\mathrm{d}x =1+ (\mu+\omega)\left(\frac{b}{c}-1+\frac{\sin(c-b)}{c}\right).
\end{equation*}
This last expression allows us to write $C_1,\,C_2$ only in terms of $b$ and $c$, and the model parameters.

\subsubsection{Coexistence region}

In order to find solutions on the coexistence region $[-b,b]$, we rewrite Eqs. \eqref{eq:steady} in more compact form
\begin{equation}
    \mathbf{\Sigma}''+M^{-1}N\mathbf{\Sigma}  = M^{-1}\mathbf{C}\,,\label{eq:steady_compact}
    \end{equation}
    where \begin{equation*}\mathbf{\Sigma}= \begin{pmatrix}
 \rho \\
 \eta \\
\end{pmatrix},\quad
    N = \begin{pmatrix}
 \mu & \omega\\
\omega & 1\\
\end{pmatrix},\quad \mathbf{C}=\begin{pmatrix}
 C_1\\
C_2\\
\end{pmatrix},
\end{equation*}
 and $M$ is defined by $\eqref{eq:M}$. With this, general solutions of \eqref{eq:steady_compact} can be written as
 \begin{equation}
       \mathbf{\Sigma}(x) = A\mathbf{v_1}e^{i\lambda_1 x}+B\mathbf{v_1}e^{-i\lambda_1 x}+D\mathbf{v_2}e^{i\lambda_2 x}+E \mathbf{v_2}e^{-i\lambda_2 x}+N^{-1}\mathbf{C}\,,\label{eq:general1}
    \end{equation}
    where $\mathbf{v_1}, \mathbf{v_2}$ are eigenvectors of $M^{-1}N$ with eigenvalues $\lambda_1^2,\lambda_2^2$,  respectively. For simplicity, we also set now
 $$\begin{pmatrix} D_1 \\ D_2
 \end{pmatrix} = N^{-1}\begin{pmatrix} C_1 \\ C_2
 \end{pmatrix}.$$ Note that both $D_1$ and $D_2$ can be written in terms of $b,c$ and the model parameters. The eigenvalues and eigenvectors of $M^{-1}N$ can be found explicitly
    \begin{equation*}
        \lambda_1^2 = \frac{\kappa+\mu-2\alpha\omega+\sqrt{\Delta}}{2\det M}\,,\qquad         \lambda_2^2 = \frac{\kappa+\mu-2\alpha\omega-\sqrt{\Delta}}{2\det M}\,;
    \end{equation*}
    \begin{equation*}
    \mathbf{v_1} = \begin{pmatrix}
 2(\alpha-\omega) \\
\mu-\kappa-\sqrt{\Delta} \\
\end{pmatrix},\qquad
\mathbf{v_2} = \begin{pmatrix}
2(\alpha-\omega) \\
\mu-\kappa+\sqrt{\Delta} \end{pmatrix};\\
 \end{equation*}
 with $$\Delta = (\mu-\kappa)^2+4(\alpha\mu-\kappa\omega)(\alpha-\omega).$$
 By looking at $\Delta$ as a quadratic polynomial in $\omega$, we see that
 $$\Delta \geq \left(1-\frac{\alpha^2}{\kappa}\right)\left(\mu-\kappa\right)^2>0,$$
 and hence $\lambda_1^2,\lambda_2^2$ are always real.

 In general, $\lambda_1^2$ is always positive and $\lambda_2^2$ can be either positive or negative. To see this, write $$\lambda_1^2=\frac{1}{2}\left(\text{tr}(M^{-1}N)+\sqrt{\text{tr}(M^{-1}N)^2-4\det M^{-1}\det N}\right).$$
 If $\text{tr}(M^{-1}N)<0$, then $\kappa+\mu-2\alpha\omega<0$ and thus
 $$\det N = \mu-\omega^2<2\alpha\omega-\omega^2-\kappa<-(\omega-\alpha)^2<0\,.$$
 Consequently $\lambda_1^2$ is always positive. However, for $\lambda_2^2$ we can write
 $$\lambda_2^2=\frac{1}{2}\left(\text{tr}(M^{-1}N)-\sqrt{\text{tr}(M^{-1}N)^2-4\det M^{-1}\det N}\right),$$ and hence $\lambda_2^2$ will be negative whenever $\text{tr}(M^{-1}N)<0$ or $\text{tr}(M^{-1}N)>0$ and $\det N<0$. In terms of $\omega$, this happens whenever $\omega>\min\left(\sqrt{\mu},(\kappa+\mu)/2\alpha\right) = \sqrt{\mu}$, where we used that $\mu>\kappa>\alpha^2$. We consider both cases separately now.

 \subsubsection{Two positive eigenvalues}

  \begin{figure}
    \centering
    \includegraphics[width = \textwidth]{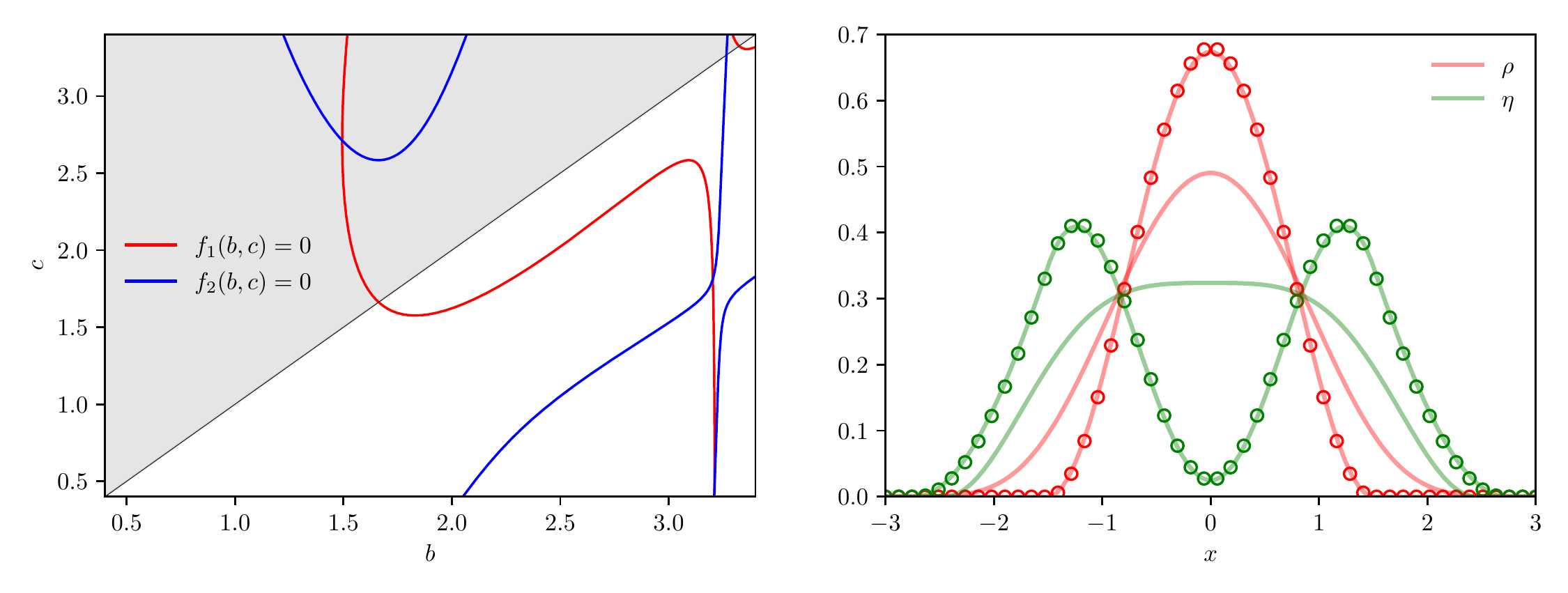}
    \caption{(left) Regularity of $\eta$ determines the support. The shaded region represents the condition $b<c$, i.e. that $\rho$ is the more cohesive population. (right) Numerical solution of the two-species model with initial condition $\rho(x,0) = \eta(x,0) = \chi_{|x|<1}/2$. Solutions are shown at $t = 1$ and $t = 25$ (solid line) and the corresponding analytical stationary solutions are also plotted (dots). The analytical and numerical stationary solutions agree perfectly. Simulation parameters: $\kappa = 2,\,\alpha = 1.3,\,\mu =4 ,\,\omega =1.8,\,m = 1,\, L = 3,\, \Delta x = 0.1,\, \Delta t = 10^{-2}$.}
    \label{fig4:steady_state_1}
\end{figure}

As discussed, this happens whenever $\omega < \sqrt{\mu}$. In this case, and using the fact that the stationary solutions are symmetric, we can write general solutions of Eq. \eqref{eq:general1} as
 \begin{subequations}
 \begin{align*}
 \rho(x) &= 2(\alpha-\omega)A_2\cos(\lambda_1 x)+2(\alpha-\omega)B_2\cos(\lambda_2 x) + D_1\,;
     \\  \eta(x) &= (\mu-\kappa-\sqrt{\Delta})A_2\cos(\lambda_1 x)+ (\mu-\kappa+\sqrt{\Delta})B_2\cos(\lambda_2 x) + D_2\,.
 \end{align*}
 \end{subequations}
 The coefficients $A_2, B_2$ can be found in terms of $b,c$ by imposing $\rho(b) = \rho'(b) = 0$. These conditions give
 \begin{subequations}
 \begin{align*}
     A_2 = -\frac{D_1\left(\lambda_1^{-1}\cot(\lambda_1 b)-\lambda_2^{-1}\cot(\lambda_2 b)\right)^{-1}}{2(\alpha-\omega)\lambda_1\sin(\lambda_1 b)}\,,\\ B_2 = \frac{D_1\left(\lambda_1^{-1}\cot(\lambda_1 b)-\lambda_2^{-1}\cot(\lambda_2 b)\right)^{-1}}{2(\alpha-\omega)\lambda_2\sin(\lambda_2 b)}\,.
 \end{align*}
  \end{subequations}
 Note that we have expressed $\rho$ and $\eta$ only in terms of the model parameters and $b,c$. In order to find these two parameters we only need to impose that $\eta$ is continuously differentiable on $x= b$. This condition gives two equations
 \begin{subequations}
 \begin{align*}
 f_1(b,c) &= \lim_{x\rightarrow b^+}\eta'(x) - \lim_{x\rightarrow b^-}\eta'(x) = 0\,;\\
  f_2(b,c) &= \lim_{x\rightarrow b^+}\eta(x) - \lim_{x\rightarrow b^-}\eta(x) = 0\,;\end{align*}
 \end{subequations} which can be solved numerically to find $b$ and $c$. This is shown in Figure \ref{fig4:steady_state_1}, alongside the corresponding stationary solutions. We see that the numerical stationary state and the solution found in this section agree perfectly.

   \subsubsection{Positive and negative eigenvalues} Conversely, whenever $\omega > \sqrt{\mu}$ we have $\lambda_2^2<0$. Using again that the stationary states are symmetric, we can write general solutions Eq. \eqref{eq:general1} as
    \begin{subequations}
 \begin{align*}
 \rho(x) &= 2(\alpha-\omega)A_2\cos(\lambda_1 x)+2(\alpha-\omega)B_2\cosh(|\lambda_2 |x) + D_1\,;
     \\  \eta(x) &= (\mu-\kappa-\sqrt{\Delta})A_2\cos(\lambda_1 x)+ (\mu-\kappa+\sqrt{\Delta})B_2\cosh(|\lambda_2| x) + D_2\,.
 \end{align*}
 \end{subequations}
Imposing  $\rho'(b) = \rho'(b) = 0$ we find
\begin{subequations}
 \begin{align*}
     A_2 = -\frac{D_1\left(\lambda_1^{-1}\cot(\lambda_1 b)+|\lambda_2|^{-1}\coth(|\lambda_2| b)\right)^{-1}}{2(\alpha-\omega)\lambda_1\sin(\lambda_1 b)}\,,\\ B_2 = -\frac{D_1\left(\lambda_1^{-1}\cot(\lambda_1 b)+|\lambda_2|^{-1}\coth(|\lambda_2| b)\right)^{-1}}{2(\alpha-\omega)|\lambda_2|\sinh(|\lambda_2| b)}\,.
 \end{align*}
\end{subequations}
And again, we have the same two conditions on the regularity of $\eta$ at $x = b$
 \begin{subequations}
 \begin{align}
 f_1(b,c) &= \lim_{x\rightarrow b^+}\eta'(x) - \lim_{x\rightarrow b^-}\eta'(x) = 0\,;\label{eq:support_1_negative}\\
  f_2(b,c) &= \lim_{x\rightarrow b^+}\eta(x) - \lim_{x\rightarrow b^-}\eta(x) = 0\,.\label{eq:support_2_negative}\end{align}\label{eq:support_negative}
\end{subequations}

\begin{figure}[t]
    \centering
    \includegraphics[width = \textwidth]{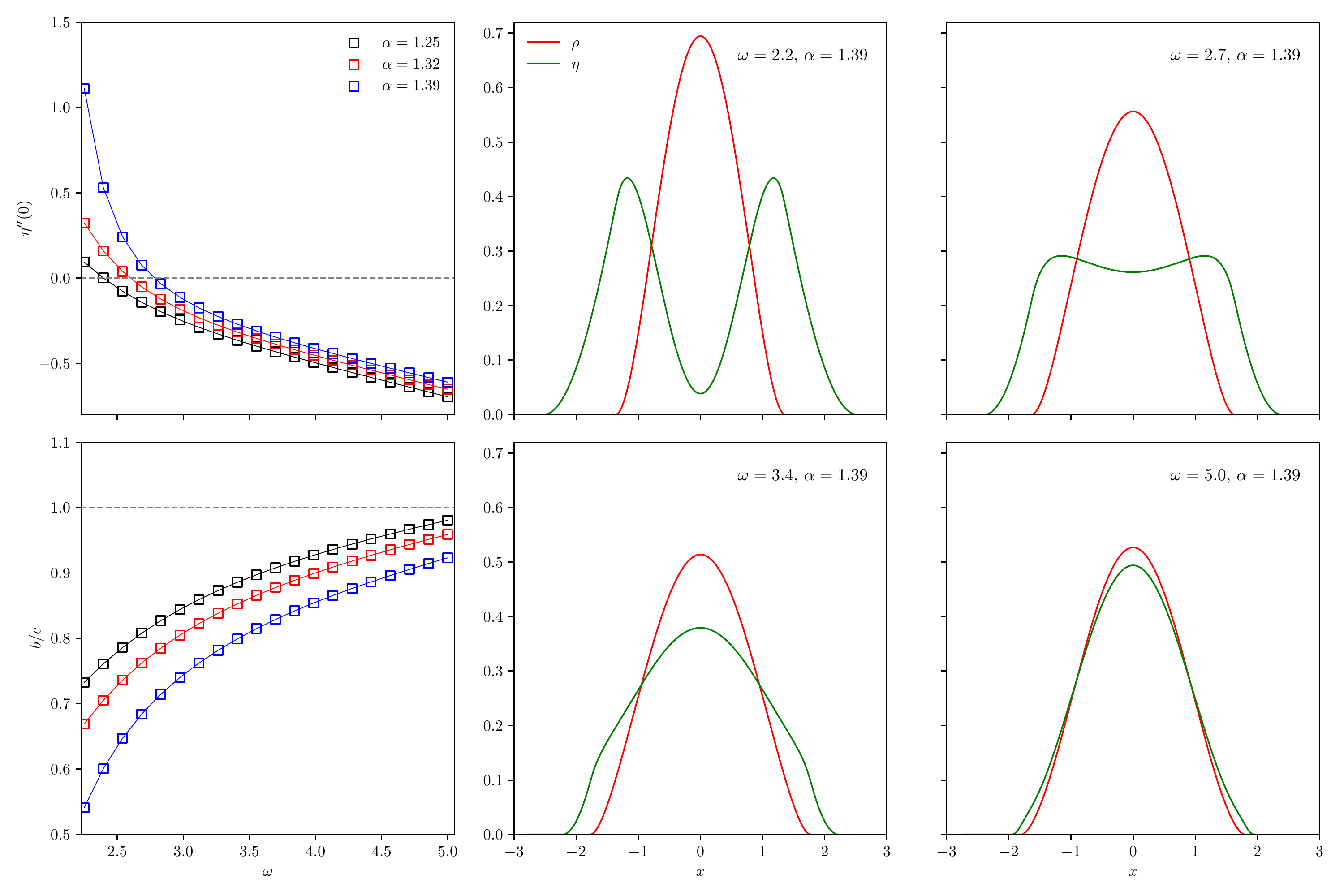}
    \caption{Engulfment-mixing transition from the analytical stationary solutions. In every case $\kappa = 2,\,\mu = 4$. On the left we plot the second derivative of the engulfing species at the origin $\eta''(0)$ and the quotient $b/c$ as a function of $\omega$ and for different values of $\alpha$. The support length relation is found by solving numerically Eqs. \eqref{eq:support_negative}. On the right, we plot different solutions, showing the transition from one pattern to the other.}
    \label{fig:mixing_engulfment}
\end{figure}

Having explicit stationary solutions is useful for predicting transitions between the different patterns shown in Figure \ref{Fig:2species_1d}. We focus here on the engulfment and mixing patterns, obtained in the strong cross-adhesion regime. Although there is no sharp transition between these two patterns, it is instructive to understand how solutions vary when increasing the cross-adhesion strength. We plot the analytical solutions in Figure \ref{fig:mixing_engulfment}. Note that in general, this is not possible for nonlocal models. One could further ask whether these calculations can provide analytical insights into the transitions between different patterns. However, the equations to determine the support of the densities ($b,c$) are complicated and need to be solved numerically. as we show in Figure \ref{fig4:steady_state_1}. In future work, it would be interesting to examine this question under a slightly simpler setting -- for instance, assuming identical self-adhesion interactions \cite{carrillo2018zoology}.

One measure to quantify when one of the populations is trapped inside the other is given by the second derivative of the engulfing species, $\eta''(0)$. This quantity is positive for engulfment and negative when the two species are mixed. Another possibility is to look at the quotient $b/c$, which should approach unity as we increase the strength of cross-adhesion between the two species. Using the expressions we found for the steady states, we plot how these quantities vary with $\omega$ and $\alpha$ in \eqref{fig:mixing_engulfment}. Again we confirm our intuition, since increasing the cross-adhesion strength -- and hence $\omega$ -- yields the expected behaviour, decreasing $\eta''(0)$ and a quotient $b/c$ that approaches unity.

 \subsection{Numerical simulations for Steinberg experiments in two dimensions}

\begin{figure}[t]
    \centering
    \includegraphics[width = \textwidth]{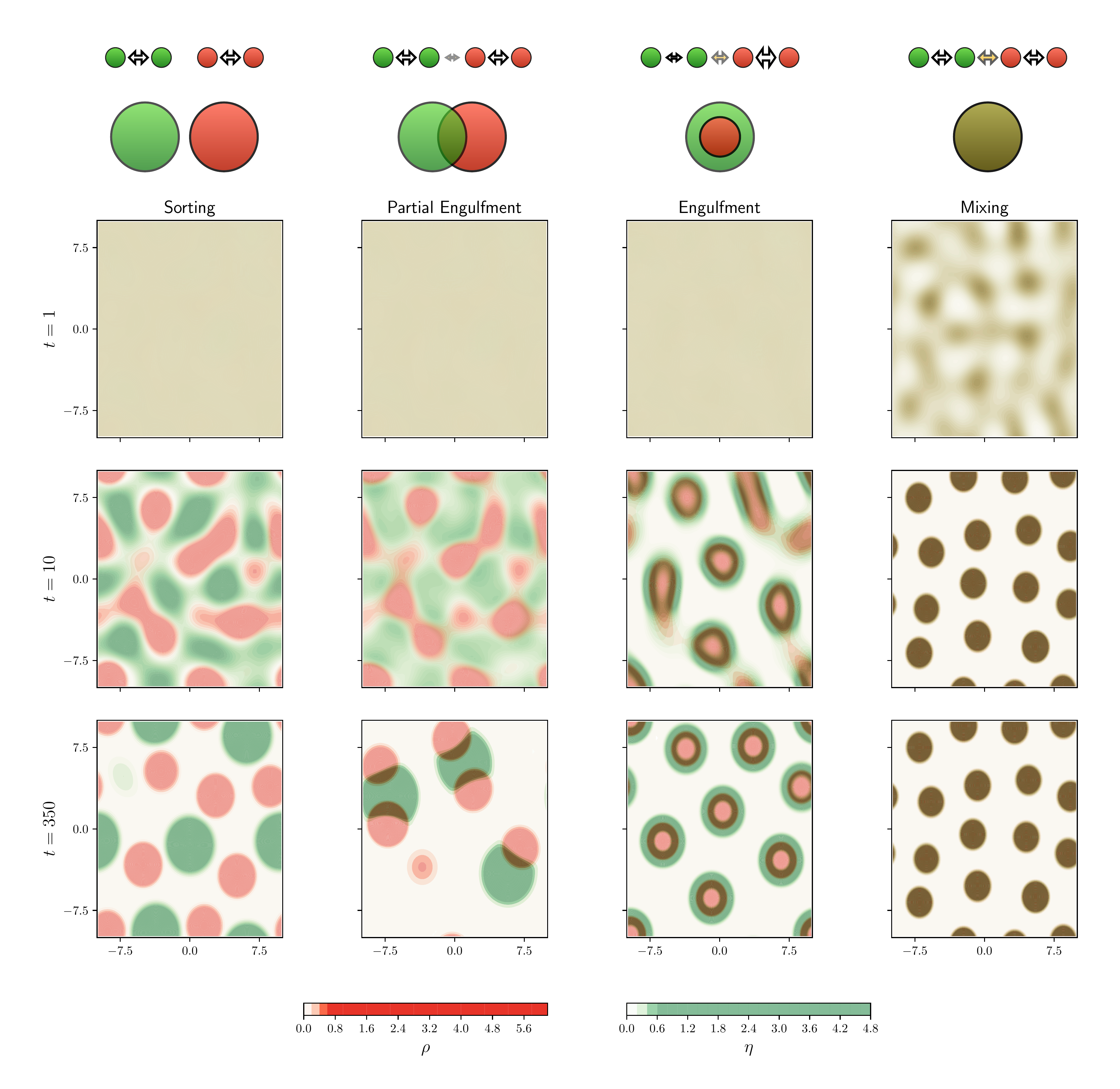}
    \caption{Numerical solutions of the local model using model parameters related to Steinberg experiments. Each column represents the solution with the same set of parameters and at different times. Mixing, $\alpha = 1.4 ,\,\omega =8$; engulfment, $\alpha =  1.3,\,\omega =2$; partial engulfment, $\alpha = 0.5 ,\,\omega = -0.02$; sorting, $\alpha = 0 ,\,\omega =-1$. In every case $\kappa = 2$ and $\mu = 4$ and also  $L = 10,\,\Delta x = 0.2,\, \Delta t =0.001$. The initial condition is the same for every experiment, $\rho(x,0), \,\eta(x,0) = 0.3$ plus a small perturbation. See \cite{figshare} for animated movies.}
    \label{fig:2D_DAH}
\end{figure}

In two spatial dimensions, the explicit calculations performed in the previous section involve Bessel functions, and imposing boundary and regularity conditions becomes a very cumbersome task. Here instead we explore the model in two dimensions numerically, performing the same type of experiments as in the one-dimensional case, which we show in Figure \ref{fig:2D_DAH}.

By choosing appropriate parameters, as explained in the previous sections, we can again recover the four patterns seen in the Steinberg experiments. The model dynamics are in general similar to the one-dimensional case, with the strong cross-adhesion regime showing a faster decay to the stationary solution. Note that although the final configurations are very different for the chosen parameters, in the weak cross-adhesion regime solutions show very similar patterns for early times. In this case, the model shows two separated timescales with solutions showing large differences only after the first one, as it already happened in the one-dimensional case (see Figure \ref{fig:energy}). In the limit of vanishing cross-adhesion we recover the typical cell sorting pattern with sharp segregation of the two species, which is only observed in models that account for population pressure. Note also that this pattern is accentuated with respect to the one-dimensional case.

\section{Conclusion and outlook}

To summarize, in this paper we have presented a local continuum model of cell-cell adhesion that takes the form of a system of thin-film equations. As discussed, the idea of describing cells and tissues using fluid-like properties has been recurrent. In fact Steinberg already thought of this analogy for developing the DAH. Note however, that the model proposed here is different from other phenomenological descriptions based on the fluid analogy \cite{AlertTrepat}, as it can be directly related to aggregation-diffusion equations. The new local model has physically interpretable parameters, and is able to explain the patterns seen in experiments and that are predicted by the DAH. To the best of our knowledge, in the continuum setting this has previously only been achieved with nonlocal models.

Of course, there are also more modern views on the DAH, such as the differential interfacial tension hypothesis \cite{BrodlandDITH,reviewCellCellAdhesion}, which is based on the idea that it is not only adhesion bonds between cells that determines tissue surface tension but also cortical tension  \cite{amack2012knowing,youssef2011quantification}. There have also been some modelling efforts to account for this, showing that when cell-cell adhesion is the dominant interaction, the DAH is successful in predicting tissue behaviour \cite{ManningFotySteinbergSchoetz}. When cortical tension is stronger, however, the DAH might not be sufficient, showing that in this regime cells cannot be considered as individual points. While this might be an interesting  point to consider, our model here builds on the adhesion-based regime, where the DAH and the particle-based approximation hold.

Our model was motivated by both experimental and theoretical studies of cell-cell adhesion. However, the same ideas can be applied in other biological contexts, where again differential surface tension between multiple species drives the formation of different patterns -- see for instance the assembly of intracellular ribonucleoprotein bodies via liquid-liquid phase separation \cite{feric2016coexisting,lu2020multiphase}. We also remark here that models describing the evolution of multicomponent liquid mixtures appear often in the context of Cahn-Hilliard equations \cite{barrett2001fully,elliott1997diffusional,mao2019phase}.

The approach taken here is based on previous studies of aggregation-diffusion systems, where the nonlocal terms are approximated by a series of terms including higher-order derivatives of the densities \cite{BernoffTopazCH,DelgadinoCH} -- note that these consider a porous-medium type repulsion with exponent three, instead of the exponent two considered here. Following the same idea, energy minimizers and linear stability for multi-species systems have also been examined \cite{EllefsenRodriguez}. All these have proven that the resulting thin-film or Cahn-Hilliard type models show interesting behaviour, but it is not yet clear how close this fourth-order approximations are, or whether one can expect the same phenomenology from nonlocal models and their local approximations.

In fact, it has already been shown that energy minimizers of the local and nonlocal models are in good agreement and have similar qualitative properties in the limit of large populations, and far from aggregation boundaries \cite{BernoffTopazCH}. In our case the choice of diffusion is different – porous-medium type with exponent two instead of three – and hence this result does not necessarily apply. However, recent work \cite{elbar2022degenerate} shows that in a similar setting, with a unit volume and compactly supported potential, the nonlocal model tends to its local approximation  in the limit where the scaling parameter tends to zero ($a\rightarrow 0$). This also raises the question of whether a similar result holds true in the case of two interacting species. In any case, we have demonstrated that the local approximation shows the same phenomena as previously used nonlocal models – as long as the used interaction potential has compact support and hence a finite interaction range. A numerical study of both models under similar conditions would also be interesting to explore.

The local model is in principle less complex and more analytically tractable than its nonlocal counterpart. Thus we believe it could offer some advantages for applications. Although solving numerically fourth order equations can be challenging, there is a simplification in the numerical scheme complexity when one approximates convolutions with local operators \cite{nonlocalNumericalSchemeRafaMarkus,nonlocalSchemceCarrilloYanghong}.
Another possibility that opens is to connect experimental data with the local model, which in its reduced form only has four parameters. This is an important reduction in contrast to having to infer interaction potentials, as is the case in the nonlocal models.

From an analytical point of view, we were able to characterize stationary solutions of the local system, which explain the patterns obtained from model parameters. We remark that this is not possible in general nonlocal models, where expressions for steady states are only available in some specific cases \cite{carrillo2018zoology}. Note however, that to obtain them, we assumed that they correspond to energy minimizers and restricted ourselves to the case of a single droplet state with compact support. We also have not quantified the dynamical properties of the system, which seems challenging but also very intriguing. While there has been intensive work studying many properties of thin-film equations, the case of two-species remains mainly unexplored. Indeed, problems like the existence of solutions for the system, added to the ones above, remain completely open.

\appendix
\section{Outline of the numerical scheme} \label{sec:numerical_scheme} We explain here the used finite-volume method, which is based on a numerical scheme for the Cahn-Hilliard equation \cite{bailo2021unconditional}. For simplicity, we only deal with the one-species case in one spatial dimension
\begin{equation*}
    \frac{\partial\rho}{\partial t} = -\frac{\partial}{\partial x}\left(\rho\,\frac{\partial}{\partial x} \left(\frac{\partial^2\rho}{\partial x^2}  + \mu^2\rho \right)\right),
\end{equation*}
subject to periodic boundary conditions.
The two-species and two-dimensional cases are a straightforward extension of the presented scheme.

The domain $[-L,L]$ is discretized into $2N+1$ equispaced cells $C_i = [x_{i-1/2},x_{i+1/2}]$ of size $\Delta x = L/N$ centered at $x_i = i\Delta x$, $i = -N,\ldots,N$. The density $\rho(x,t)$ is approximated by the cell average $\rho_i(t)$, which is defined as
\begin{equation*}
    \rho_i(t) = \frac{1}{\Delta x}\int_{C_i} \rho(x,t) \,\mathrm{d}x.
\end{equation*}

Integrating over a test cell $C_i$, we obtain the following system of ODEs for $\rho_i$
\begin{equation}
    \frac{\mathrm{d}\rho_i(t)}{\mathrm{d}t} = -\frac{F_{i + 1/2}(t)-F_{i-1/2}(t)}{\Delta x},\label{eq:ODEs_numerical}
\end{equation}
where the numerical flux $F_{i + 1/2}(t)$ is an approximation of the flux in the PDE given by: $-\rho v:=\rho\left(\rho_{xx}+\mu^2\rho\right)_x$. In order to construct this approximation, we first define the discrete velocities $v_{i+1/2}$ as
\begin{equation*}
    v_{i+1/2} = -\frac{\xi_{i + 1}-\xi_i}{\Delta x},\quad \xi_i = \left(\Delta\rho\right)_i + \mu^2\rho_i,
\end{equation*}
where $\left(\Delta\rho\right)_i$ is the usual one-dimensional second-order approximation of the Laplacian
\begin{equation*}
    \left(\Delta\rho\right)_i = \frac{\rho_{i+1}-2\rho_i+\rho_{i-1}}{\Delta x^2}.
\end{equation*}
Then we follow an upwind approach to calculate the fluxes
\begin{equation*}
    F_{i + 1/2} = (v_{i+1/2})^+(\rho_{i+1})^+ + (v_{i+1/2})^-(\rho_{i})^+,
\end{equation*}
with
\begin{equation*}
    (v_{i+1/2})^+ = \max\left(v_{i+1/2},0\right),\quad (v_{i+1/2})^- = \min\left(v_{i+1/2},0\right).
\end{equation*}

With this construction we only need to solve the system of ordinary differential equations given by Eq. \eqref{eq:ODEs_numerical}. In our case, these are solved using a fourth-order Runge-Kutta method. The chosen time and space discretizations are specified on the caption of each figure. We finally remark that we do an explicit in time discretization of Eq. \eqref{eq:ODEs_numerical}. As proven in \cite{bailo2021unconditional}, this explicit discretization preserves positivity under a CFL condition, while the energy decay is only kept at the semidiscrete level in Eq. \eqref{eq:ODEs_numerical}.

\end{document}